\documentclass[preprint]{aastex}
\usepackage{graphics,wrapfig,amsmath,color,rotating}
\begin{document}

\title{Collisional Evolution of Ultra-Wide Trans-Neptunian Binaries}
\author{\bf Alex H. Parker$^{1,2}$ \& JJ Kavelaars$^3$}
\email{alexhp@uvic.ca}
\affil{\emph{ $^{1}$Department of Astronomy, University of Victoria,  $^{2}$Harvard-Smithsonian Center for Astrophysics, $^{3}$Herzberg Institute of Astrophysics, National Research Council of Canada}}

\shortauthors{Parker et al.}

\begin{abstract}

The widely-separated, near-equal mass binaries hosted by the cold Classical Kuiper Belt are delicately bound and subject to disruption by many perturbing processes. We use analytical arguments and numerical simulations to determine their collisional lifetimes given various impactor size distributions, and include the effects of mass-loss and multiple impacts over the lifetime of each system. These collisional lifetimes constrain the population of small ($R \gtrsim 1$ km) objects currently residing in the Kuiper Belt, and confirm that the size distribution slope at small size cannot be excessively steep --- likely $q \lesssim 3.5$. We track mutual semi-major axis, inclination, and eccentricity evolution through our simulations, and show that it is unlikely that the wide binary population represents an evolved tail of the primordially-tight binary population. We find that if the wide binaries are a collisionally-eroded population, their primordial mutual orbit planes must have preferred to lie in the plane of the solar system. Finally, we find that current limits on the size distribution at small radii remain high enough that the prospect of detecting dust-producing collisions in real-time in the Kuiper Belt with future optical surveys is feasible.

\end{abstract}


\maketitle

\section{Introduction}

The dynamically-cold component of the classical Kuiper Belt is host to a very high fraction of binary systems, and some of these systems have very wide separations. These ultra-wide Trans Neptunian Binaries (TNBs) have been shown to be extremely delicate, sensitive to collisional disruption (Petit \& Mousis 2004, Nesvorn\'{y} et al. 2011, Parker et al. 2011) and disruption by close encounters with Neptune (Parker \& Kavelaars 2010). 

Given the characterized sample of ultra-wide TNBs presented in Parker et al. (2011), we seek to accurately determine the collisional lifetimes of these ultra-wide systems and their implications for the current state of the Kuiper Belt. To do this, we expand upon the analytical estimates of Petit \& Mousis (2004) in \S2, then desribe numerical simulations we performed to more accurately account for the effects of mutual eccentricity, mass loss, and multiple impactors in \S3. We find a simple empirical correction to the analytic predictions described in \S2 which accurately reproduces the collisional lifetimes determined by our simulations.

Armed with these new estimates of collisional lifetimes under a variety of assumed impactor populations, in \S4 we determine the properties of these impactor populations (representing the population of $\gtrsim1$ km) that are allowed given conservative assumptions about the primordial fraction of ultra-wide TNBs in the classical Kuiper Belt. This impactor population is extremely difficult to constrain observationally; TNOs with $R \sim 1$ km are unlikely to ever be detected in reflected light (with magnitudes $\gtrsim33$), and at present the only limits that exist on their numbers are based on searches for stellar occultations (Schlichting et al. 2009, Wang et al. 2009, Bickerton, Kavelaars \& Welch 2008). Determining the behavior of the TNO size distribution at small sizes is critical for understanding the accretion and collisional history of the outer solar system.

In \S5 we discuss the evolution of the properties of the binary mutual orbits over their lifetimes as they are subjected to collisions, and discuss the effects of these collisions on the interpretation of the current orbital distributions. In general we find that it is unlikely for the ultra-wide binaries to have evolved from initially tighter orbits, that their primordial mutual inclination distribution must have been even colder than it is currently, and that their current roughly equal numbers of prograde and retrograde orientations likely reflects the primordial distribution.

In the following discussions we will adopt the nomenclature of Parker et al. (2011) for discussing binary populations and separations, with ``tight'' TNBs being those with mutual semi-major axes significantly less than 5\% of their Hill radius, ``wide'' TNBs with mutual semi-major axes of approximately 5\% of their Hill radius, and ``ultra-wide'' TNBs being those with mutual semi-major axes significantly exceeding 5\% of their Hill radius.

We conclude with a discussion of the implications of the existence of a single low-mass, widely separated system like 2000 CF$_{105}$ and the importance of identifying the prevalence of systems like it in the current Kuiper Belt, implications of post-formation collisional evolution for our understanding of the binary formation mechanism(s), and also discuss the probability and utility of detecting transient brightening events caused by catastrophic collisions like those simulated in this work in future surveys like LSST.

\section{ Analytical Estimates of Collisional Lifetimes }

Following Petit \& Mousis (2004), the radius $R_i$ of an impactor onto the secondary which can unbind (i.e., cause total system energy to exceed zero) a binary can be approximated by

\begin{equation}\label{R_est}
R_i \simeq R_s \left( \frac{ 0.62 }{ V_i} \right)^{\frac{1}{3}} \left( \frac{G M_{sys} }{a_m} \right)^{\frac{1}{6}},
\end{equation}

\noindent where $R_s$ is the radius of the system's secondary, $V_i$ is the velocity of the impactor, $M_{sys}$ is the total mass of both binary components, and $a_m$ is the mutual semi-major axis of the binary. For the ultra-wide TNBs presented here, collisions at typical Kuiper Belt relative velocities of $V_i \sim 1$ km s$^{-1}$, Eqn. \ref{R_est} estimates that the required impactor radii range from $\sim$1.6 km (for 2000 CF$_{105}$) to $\sim$5.7 km (for 2001 QW$_{322}$).

We can estimate the mean time between single impacts large enough to unbind the binary,

\begin{equation}
\bar{t} \simeq \left( P_{i} \, R_s^2 \, N( R > R_i ) \right)^{-1},
\end{equation}

\noindent where $P_{i}$ is the intrinsic collision probability in km$^{-2}$ yr$^{-1}$ for objects in the Classical Kuiper Belt, and we nominally adopt the value of $1.3\times10^{-21}$ (Farinella et al. 2000). This intrinsic collisional probability is determined by the orbital distribution of the population, and several estimates exist in literature which use different approaches (particle-in-a-box vs. analytical estimates) and assume different intrinsic orbital distributions. Later we explore the effects of adopting a different value for $P_i$. To simplify, we do not include the radius of the impacting objects in the estimate of the collisional cross-section, which results in an estimate of ``head on'' collisions instead of grazing collisions. 

Given a power-law size distribution of the form $N(> R) = N_0 (R / R_0)^{1-q}$, we find

\begin{equation}\label{tau_single}
\bar{t} \simeq \left( P_{i} \, R^2_s \, N_0 (R_i / R_0)^{1-q} \right)^{-1} =  \left( P_{i} N_0 R_0^{q-1} R^{3-q}_s \left( \frac{ 0.62 }{ V_i} \right)^{\frac{1-q}{3}} \left( \frac{G M_{sys}  }{a_m} \right)^{\frac{1-q}{6}} \right)^{-1}
\end{equation}

Noting that collisions onto the primary of a binary system can also unbind the system, our analytic estimate of the mean system lifetime is half the harmonic mean of the average time between unbinding collisions for the primary and secondary components,

\begin{equation}\label{tau_analytic}
\tau_a \simeq \left( P_{i} N_0 R_0^{q-1} \left( R_s^{3-q} + R_p^{3-q} \right) \left( \frac{ 0.62 }{ V_i} \right)^{\frac{1-q}{3}} \left( \frac{G M_{sys}  }{a_m} \right)^{\frac{1-q}{6}} \right)^{-1}.
\end{equation}

For a system with nearly-equal mass components, adding this second decay channel reduces the mean lifetime by up to a factor of two.

\subsection{Lifetime ratios: Separation and mass effects}

In order to determine the relative importance of initial binary separation vs. binary mass with respect to survival time, we consider the lifetime estimate Eqn. \ref{tau_analytic} in the case where the binary has equal-massed components with equal bulk densities --- thus, $R_p = R_s$ and $M_s = M_p  = M_{sys}/2$. The ratio of lifetimes of two binaries immersed in the same collisional environment is:

\begin{equation}\label{tau_ratio_aRH}
\tau_1 / \tau_2 = \left( \frac{ R_{s2} }{ R_{s1} } \right)^{3-q}   \left(   \frac{  a_{m1} M_{sys2} } { a_{m2} M_{sys1} } \right)^{\frac{1-q}{6} } =
\left( \frac{ M_{sys1} }{ M_{sys2} } \right)^{\frac{2}{9} (2q - 5) }   \left(   \frac{  a_{m1} / R_{H1} } { a_{m2} / R_{H2} } \right)^{\frac{1}{6}(1-q) },
\end{equation}

\noindent where subscripts indicate the properties of the first or second binary system being compared, and $R_H$ is the is the Hill radius of a given system given by $R_H = a_{out} \left( \frac{M_{sys} }{3 M_{\odot}} \right)^{\frac{1}{3}} $ where $a_{out}$ is the Heliocentric semi-major axis of the binary system's barycenter. This ratio of lifetimes allows us to compare the importance of binding ($a_m / R_H$) to system mass in a given impactor regime. The indices on the two terms sum to zero when $q= 17/5 = 3.4$. At this slope, if system 1 has half the mass of system 2, then it will have to have half the $a_m /R_H$ separation of system 2 in order to have the same lifetime. At steeper slopes, mass becomes the weakly dominant term, while for shallower slopes $a_m /R_H$ has the largest effect on the lifetime.

For $q=2.5$, $\tau_1 / \tau_2$ is independent of the ratio of system mass, and for shallower slopes $\tau_1 / \tau_2$ is actually inversely correlated with the ratio of system mass --- an increased system mass leads to a \textit{decreased} lifetime. This is a consequence of the impactor spectrum becoming very flat, resulting in the dominant mass effect in a system's lifetime becoming its cross section for collisions (which increases with system mass).

To illustrate, it is interesting to compare two systems with roughly similar values of $a_m /R_H$ but significantly different masses. 2003 UN$_{284}$ and 2000 CF$_{105}$ are a good comparison; both are very widely separated, and the ratio of their best-fit $a_m /R_H$ values is $\sim0.86$. However, the ratio of their system masses is relatively high at $\sim6.8$. Both have comparable mass ratios ($\sim3$). At a collisional equilibrium slope of $q=3.5$ we would expect a ratio of lifetimes of approximately $\tau_{UN284} / \tau_{CF105} \sim 2.5$. For $q=2.5$ we would expect them to have roughly equal lifetimes, while for an even shallower slope $q=2$ (comparable to that currently measured for small objects, eg. Fraser \& Kavelaars 2009) we would expect 2003 UN$_{284}$ to actually have a shorter lifetime than the much less massive and slightly more widely separated 2000 CF$_{105}$, with $\tau_{UN284} / \tau_{CF105} \sim 0.68$.

\section{ Numerical Simulations}

The analytical estimates presented in the previous section do not account for a number of important effects, including mass loss, orbital evolution through multiple impacts, and the eccentricity of the orbit. To more accurately determine the collisional lifetimes of these binary systems, we performed a series collisional bath simulations for each system. We subject each binary to a series of impulses, corresponding to collisions with impactors drawn from a realistic size distribution, and monitor the stability of the binary after these encounters.

Our simulated impactor population was sampled from a size distribution characterized by a power law normalized at $N( R > 1$ km$)$. We chose this radius as it was close to the impactor size required to disrupt these binary systems according to our analytical estimates. This size distribution has a slope extending to larger and smaller sizes $q_{\mbox{small}}$, and at $R=31$ km we break to a large-object slope similar to that measured in the Kuiper Belt, with $q_{\mbox{large}}=4.8$ (eg., Fraser \& Kavelaars 2009). We extrapolate the small-object slope down to a minimum impactor size of $R_{\mbox{min}} = 200$ m. 

Given the total number of impactors considered, $N(R > R_{\mbox{min}})$, we then estimated the average time between collision events:

\begin{equation*}
\bar{t} = \left( P_{i} \, \sigma \, N(R > R_{\mbox{min}}) \right)^{-1},
\end{equation*}

\noindent where $P_{i}$ is the intrinsic collision probability in km$^{-2}$ yr$^{-1}$ for objects in the Classical Kuiper Belt, and we nominally adopt the value of $1.3\times10^{-21}$ (Farinella et al. 2000), while $\sigma$ is the collisional cross section (without any $\pi$ term). This cross section is adjusted to account for the radii of the impacting population in the following way: we inflate the radii of the binary components by a fixed buffer size of four times the primary radius, $\sigma = (5R_{p})^2 + (R_{s} + 4R_{p})^2$. At each iteration, we add $\bar{t}$ to the total elapsed time, then sample an impactor radius $R_i$ from the impactor size distribution. We then determine if a near-collision occurs with the primary or the secondary, weighting the relative probability by the relative collision cross section of each component (including the extra buffer radius of four primary radii), and generate an impact parameter $b$ drawn uniformly over the area of the component which was selected. If $b \leq R_0 + 0.9R_i$ (where $R_0$ is the radius of the binary component potentially suffering the collision without the added buffer radius) a collision is taken to actually occur. In general, radii and mass are related by assuming a bulk density $\rho=1$ g cm$^{-3}$ unless otherwise stated.

If a collision occurs, we then randomly generate the binary's mean anomaly $M$ to determine the orbital phase of the binary system. The relative position $\vec{x}$ and velocity $\vec{v}$ of the binary components are estimated, and an impact trajectory is generated from a uniform sphere. All collisions are assumed to occur with relative velocity of $V_i = 1$ km s$^{-1}$ unless otherwise stated.

To improve the realism of these simulations, we also treat the mass-loss during collisions, using the strength laws found by Benz \& Asphaug (1999) for ice in impacts at velocities on the order of 0.5---3 km s$^{-1}$. Based on the kinetic energy of the impactor $K\!E_i$, we estimate the mass of the largest remaining fragment given the relationship

\begin{equation}\label{M_lrf}
\gamma \equiv \frac{M_{\mbox{lrf}}}{M_{0}} = 1 - 0.5\left( \frac{K\!E_i}{M_0 Q_D^* } \right),
\end{equation}

\noindent where $M_0$ is the mass of the parent body, and $Q_D^*$ is the specific energy required to disrupt 50\% of the mass of the parent body, given by

\begin{equation}\label{Q_star}
Q_D^* = 7\times10^7 \left( \frac{R_0}{\mbox{1 cm}} \right)^{-0.45} + 2.1 \rho \left(   \frac{R_0}{\mbox{1 cm}} \right)^{1.19} \mbox {erg gram}^{-1},
\end{equation}

\noindent where $\rho$ is the density of the parent body in gram cm$^{-1}$.

For ``small'' collisions ($\gamma \gtrsim 0.8$), we assume momentum is conserved with perfectly inelastic collisions (with all of the momentum of the impactor translated into the largest remaining fragment), ie. $\vec{v}_{b,1} = \vec{v}_{b,0} + \vec{v_{i}}\frac{M_i}{M_{lrf}}$, where subscripts $b$ and $i$ indicate binary components and impactor, respectively. We treated all velocity changes as velocity changes of the secondary with respect to the primary - therefore if a collision occurs on the primary, 

\begin{equation}
\vec{v}_{\mbox{secondary},1} = \vec{v}_{\mbox{secondary},0} - \vec{v_{i}}\frac{M_i}{\gamma M_{\mbox{primary}}},
\end{equation}

whereas if a collision occurs on the secondary,

\begin{equation}
\vec{v}_{\mbox{secondary},1} = \vec{v}_{\mbox{secondary},0} + \vec{v_{i}}\frac{M_i}{\gamma M_{\mbox{secondary}}}.
\end{equation}

\begin{figure}
\begin{centering}
\includegraphics[width=0.5\textwidth]{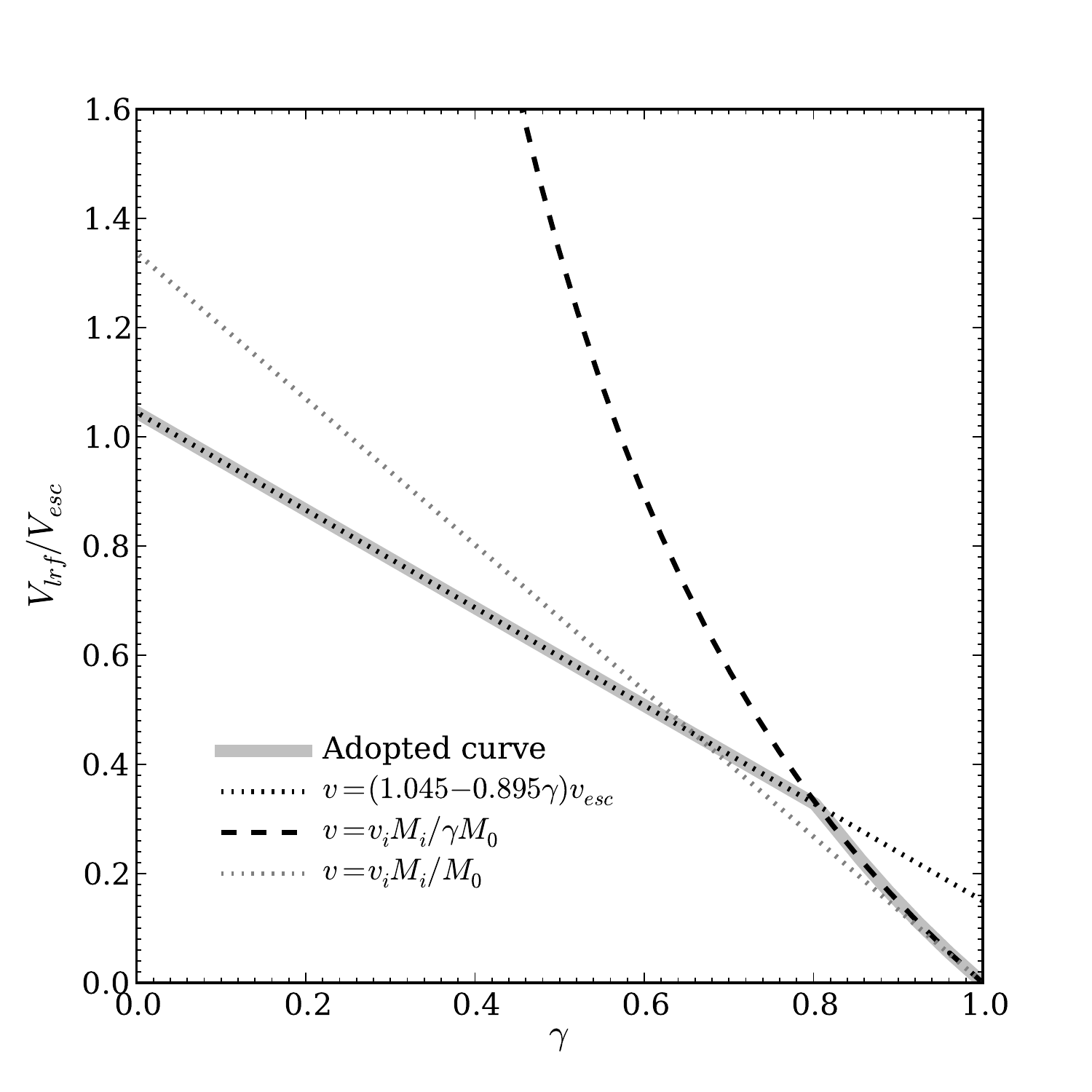}
\caption{Ratio of velocity of largest remaining fragment $V_{lrf}$ to parent body's escape velocity $V_{esc}$ for a collision which results in a mass ratio of the largest remaining fragment to the parent body of $\gamma = M_{lrf} / M_0$. Curves illustrate different schemes for estimating $V_{lrf} / V_{esc}$; adopted curve should be compared to results of numerical simulations by Benz \& Asphaug (1999, see Figures 15 \& 16).}
\label{gamma_v}
\end{centering}
\end{figure}

In larger collisions where a significant amount of mass loss occurs, much of impactor's momentum is translated into small fragments, and the largest remaining fragment experiences a much smaller change in velocity. We use a simple piecewise-linear prescription with $\gamma$ to approximately reproduce the velocity of the largest remaining fragment $V_{\mbox{lrf}}$ found by Benz \& Asphaug (1999), given by

\begin{equation}
V_{\mbox{lrf}} = \mbox{min}\left(   V',   (1.045 -0.895\gamma)V_{\mbox{esc}} \right),
\end{equation}

\noindent where $V'$ is the velocity that would be expected if all of the momentum of the impactor was translated into the largest remaining fragment, and $V_{\mbox{esc}}$ is the escape velocity of the parent body. Figure \ref{gamma_v} illustrates this velocity distribution and compares it to other schemes for treating momentum conservation for massive collisions.

After each collision, the cross-section of each component is re-computed given their new mass, and the average time between all future collisions $\bar{t}$ is re-calculated to reflect the change in collisional cross section. The total time assumed to have elapsed at this point is the sum of all the $\bar{t}$ values between all impacts preceding the latest impact.

Given a new velocity vector $\vec{v}_{1}$ post-impact, we transform coordinates to ($a,e,i$) space. If the system has become unbound, if the mutual apocenter has grown larger than the system's Hill radius $R_{H}$, or if the components have merged (mutual pericenter drops to less than the tidal Roche limit), integration is stopped. The radius of the final impactor that disrupted the system, the survival time $\tau$ (taken to be the elapsed time), and the total mass lost by each component over its lifetime is recorded.

\begin{figure*}
\begin{centering}
\includegraphics[width=0.45\textwidth]{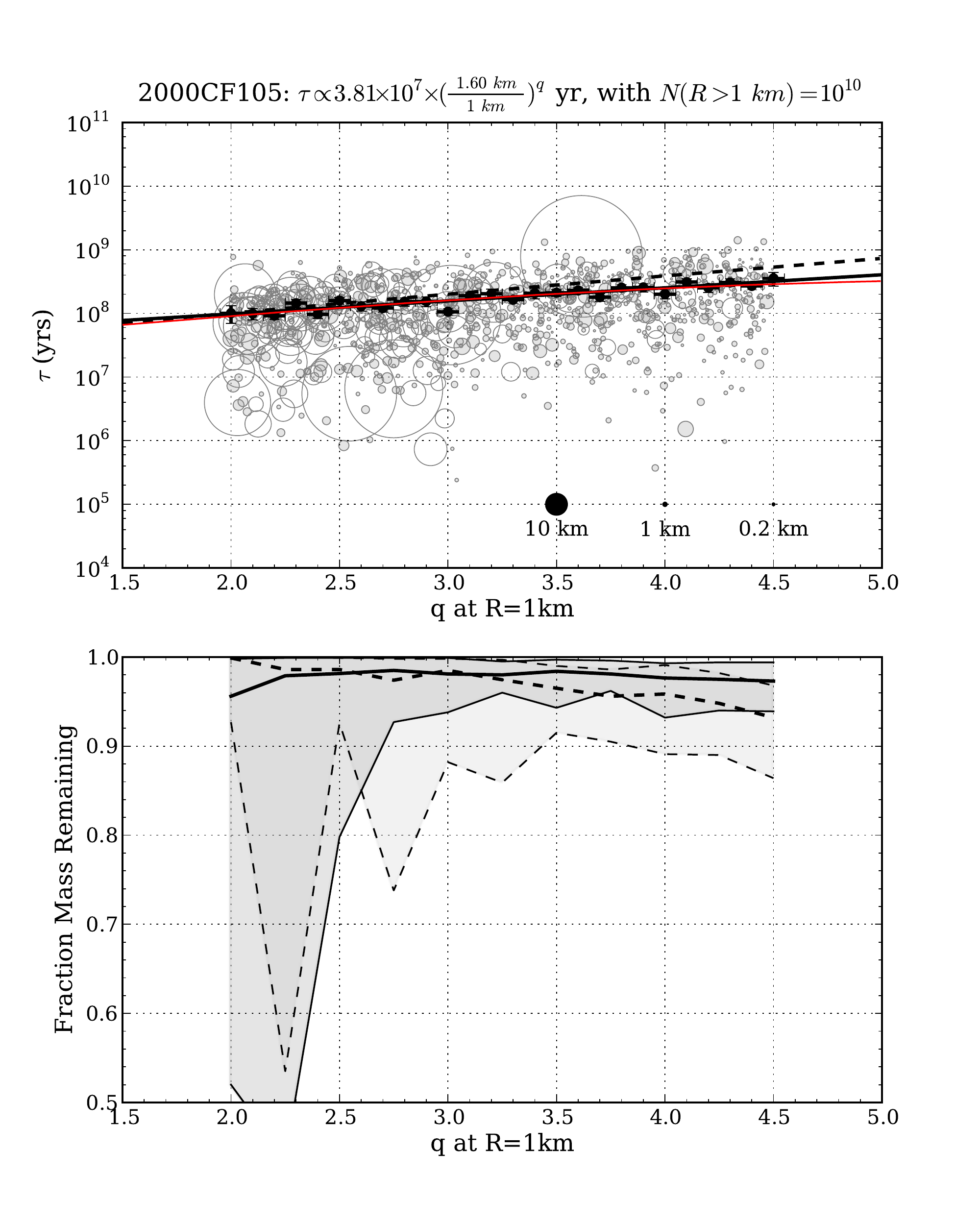}
\includegraphics[width=0.45\textwidth]{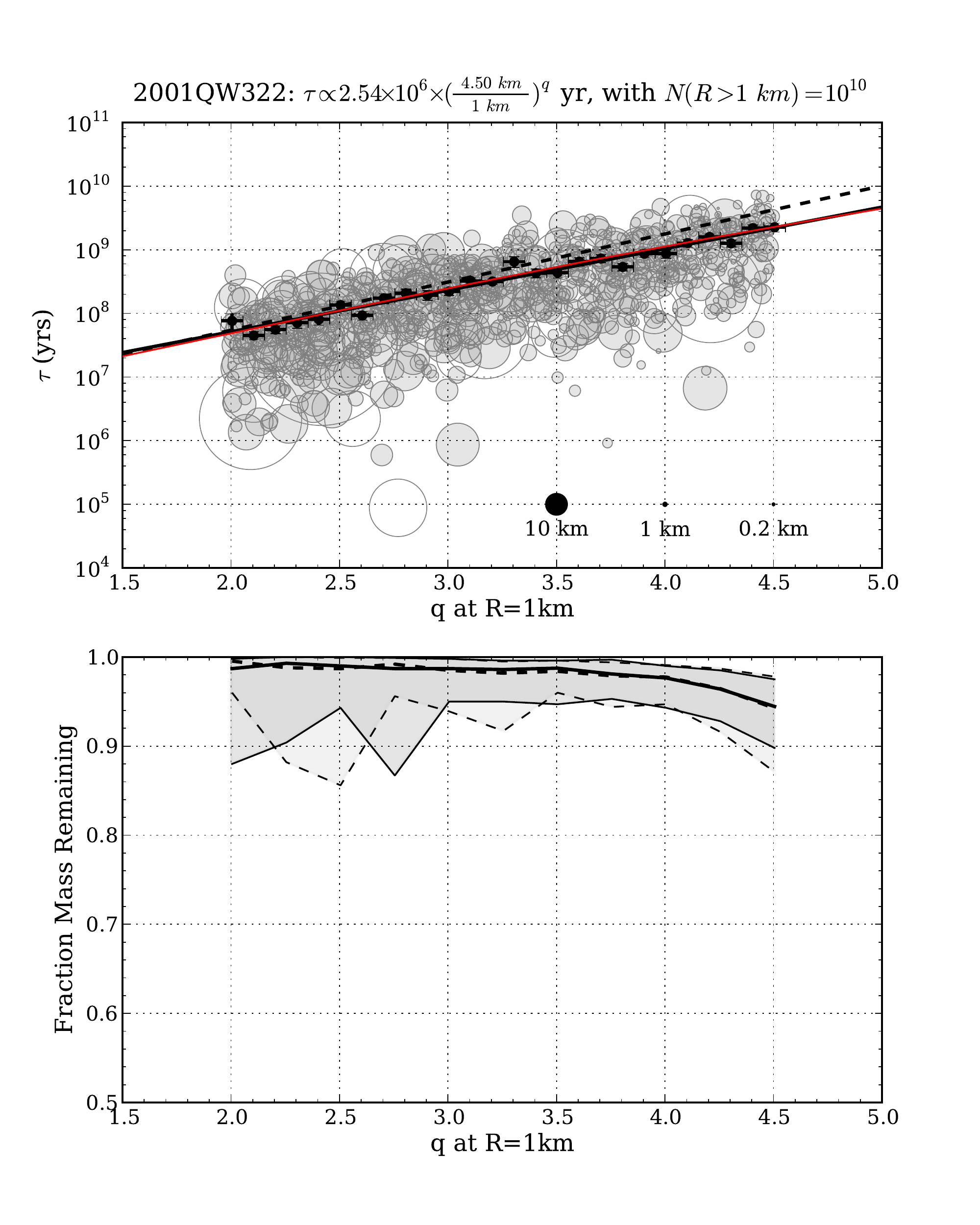}
\caption{Results from 1000 collisional bath simulations for the binary systems 2000 CF$_{105}$ (left panels) and 2001 QW$_{322}$ (right panels), where $N\!(R\!>\!1\mbox{ km})$ is held fixed at $1\times10^{10}$ and differential size distribution power-law slope $q$ is varied between 2.0 and 4.5. Top panels: Points represent time until system was disrupted in each case, while radius of each point represents the radius of the impactor that caused disruption, and red line shows fit to mean system lifetime as a function of $q$. Open points illustrate a system where one or both components lost more than 50\% of its mass by the end of the system lifetime. Bottom panels: average fraction of mass remaining in primary (heavy solid line) and secondary (heavy dashed line), and standard deviations for each (shaded regions). These two systems represent the extreme values of the critical impactor radius $r_c$ whose space density determines system lifetime.}
\label{QW}
\end{centering}
\end{figure*}

Figure \ref{QW} illustrates the result of integrating 1000 realizations of two binary systems (2000 CF$_{105}$ and 2001 QW$_{322}$), with $N(R > 1$ km$)$ held fixed and $q_{\mbox{small}}$ randomly selected at each realization from $ 2.0 < q_{\mbox{small}} < 4.5 $. For steeper size distribution slopes, smaller objects will often cause the final disruption of a system. These two systems were selected for illustration because the represent the extremes of $r_c$ for all the systems characterized.

For most realizations, the total mass loss suffered by the system is less than 10\% of its initial mass, confirming the result found by Petit \& Mousis (2004) which found that for reasonable size distributions, shattering collisions are much less important for disrupting these binary systems compared to smaller perturbations. The average mass lost by the two systems 2000 CF$_{105}$ and 2001 QW$_{322}$ is illustrated in Figure \ref{QW} as well, and generally two trends can be seen; these trends are a result of the way the impactor size distribution is normalized. At low $q_{\mbox{small}}$, relatively large amounts of mass can be lost to single (large) impactors, but as $q_{\mbox{small}}$ increases the number of large impactors decreases (since we hold $N(R> 1$ km) fixed) so the total amount of mass lost decreases with increasing $q_{\mbox{small}}$. However, for large $q_{\mbox{small}}$, the cumulative effect of many small impactors begins to matter, and as $q_{\mbox{small}}$ increases the number of these small impactors increases, and more mass is lost at higher $q_{\mbox{small}}$.

These simulations are similar to those run by Nesvorn\'{y} et al. (2011), but with several key differences. Here, we treat binaries of arbitrary mass ratio, where Nesvorn\'{y} et al. (2011) treat only those with initial mass ratios of unity. The simulations of Nesvorn\'{y} et al. (2011) immerse the binaries in a self-consistently evolving impactor distribution, so that the size distribution of impactors changes with time, whereas the shape of the size distribution in our model is assumed to be fixed in time --- however, uniform dynamical depletion of the impactor population can be accounted for in a trivial way with our model. This is discussed in Section \ref{deplete}.

\subsection{Interpretation of simulation results}

As a consequence of the way the impactor size distribution is normalized at the number with $R>1$ km, the trend in system lifetime $\tau$ vs. $q_{\mbox{small}}$ indicates the characteristic radius $r_c$ of an impactor whose space density is the critical factor in determining the lifetime of a given binary. For a system with $r_{c} \sim 1$ km, the trend in the survival time vs. $q_{\mbox{small}}$ should be approximately flat (as seen for 2000 CF$_{105}$, with $r_c = 1.65$ km), whereas if $r_{c} > 1$ km the mean survival time should increase with $q_{\mbox{small}}$ (as seen for 2001 QW$_{322}$, with $r_c = 4.35$ km) as long as $N (r > 1$ km$)$ is held fixed. 

We fit the the mean lifetime $\bar{\tau}$ as a function of slope $q$ given the collisional circumstances used in the simulation,
\begin{equation}\label{tau0}
\bar{\tau}(q) = K\times \left( \frac{ r_c \mbox{ km}} { 1 \mbox{ km} } \right)^{q}.
\end{equation}

The parameters $K$ and $r_{c}$ for each system are given in Table 1. Comparing the values of $r_c$ to the values of $r_i$ predicted by Eqn. \ref{R_est}, we see that they are generally very similar. 

The binned mean lifetimes can also be well reproduced by applying a simple correction to the analytical lifetime estimates from the previous section. Given that the analytical lifetime estimate underestimates the effect of multiple small impactors (both in random walk of orbital elements and mass loss through many erosive collisions), we might expect this correction to go as some power-law with index $\sim q$. We find that the ratio between the analytic lifetime estimate $\tau_a$ and the numerical simulation estimate $\tau_{sim}$ is well reproduced for all binaries by the function

\begin{equation}\label{fit_corr}
f = \tau_{a} / \tau_{sim} = 0.007 \times 3.12^{q} + 1
\end{equation}

\begin{figure}
\begin{centering}
\includegraphics[width=1\textwidth]{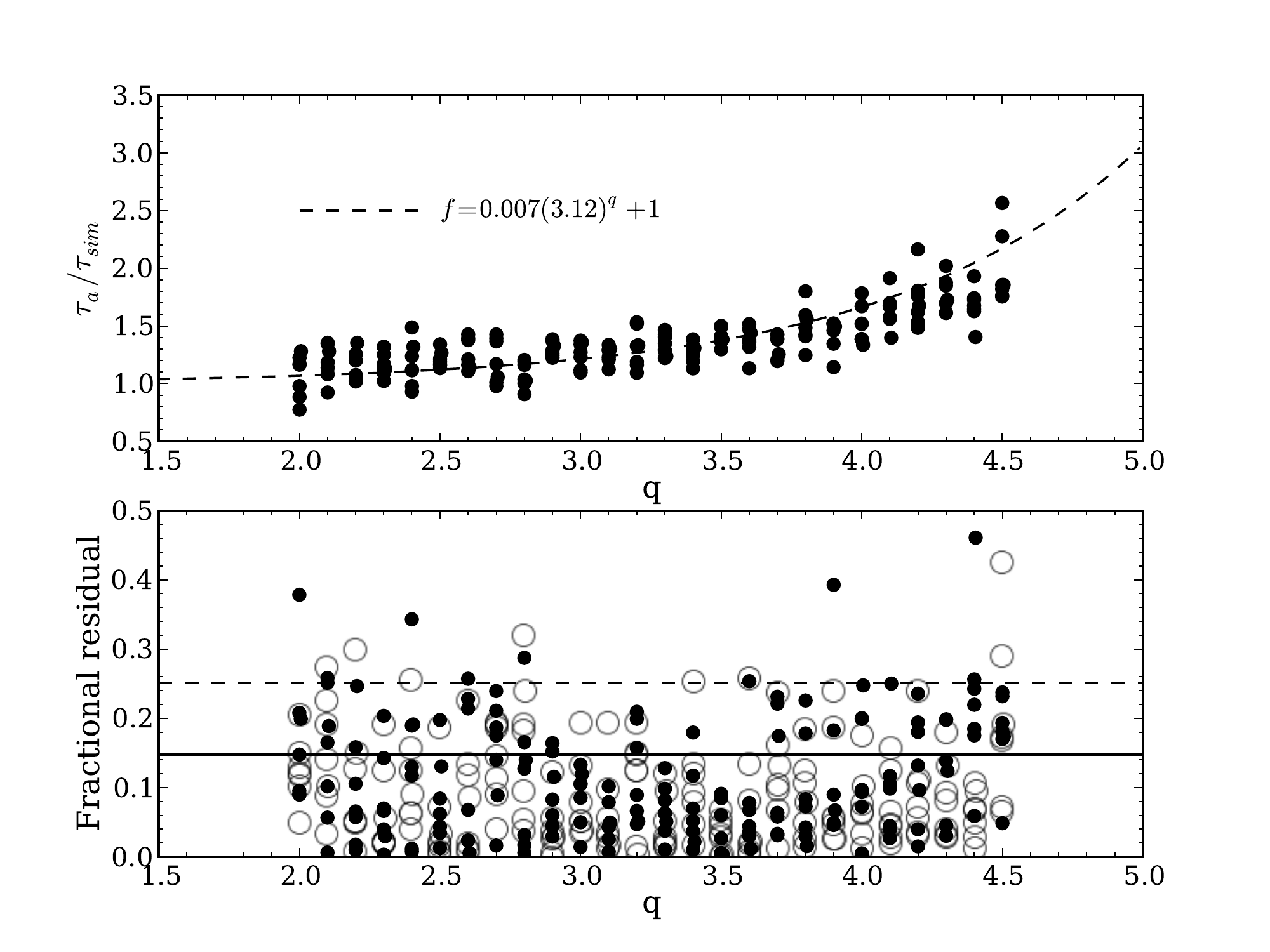}
\caption{Top panel: Ratio of analytical estimate of system lifetime $\tau_a$ to binned results of collisional simulations $\tau_{sim}$. Also illustrated is the correcting function $f$. Bottom panel: Fractional residuals between corrected lifetime and binned simulation results (filled points) and between power-law fit to simulation results and the binned simulation results (open points). Note that, in general, the corrected analytical estimates have a comparable or lower scatter than the power-law fit. 68\% and 95\% contours of fractional residuals between corrected lifetime and binned simulation results shown by solid and dashed line, respectively. }
\label{tau_corr}
\end{centering}
\end{figure}

\begin{figure}
\begin{centering}
\includegraphics[width=1\textwidth]{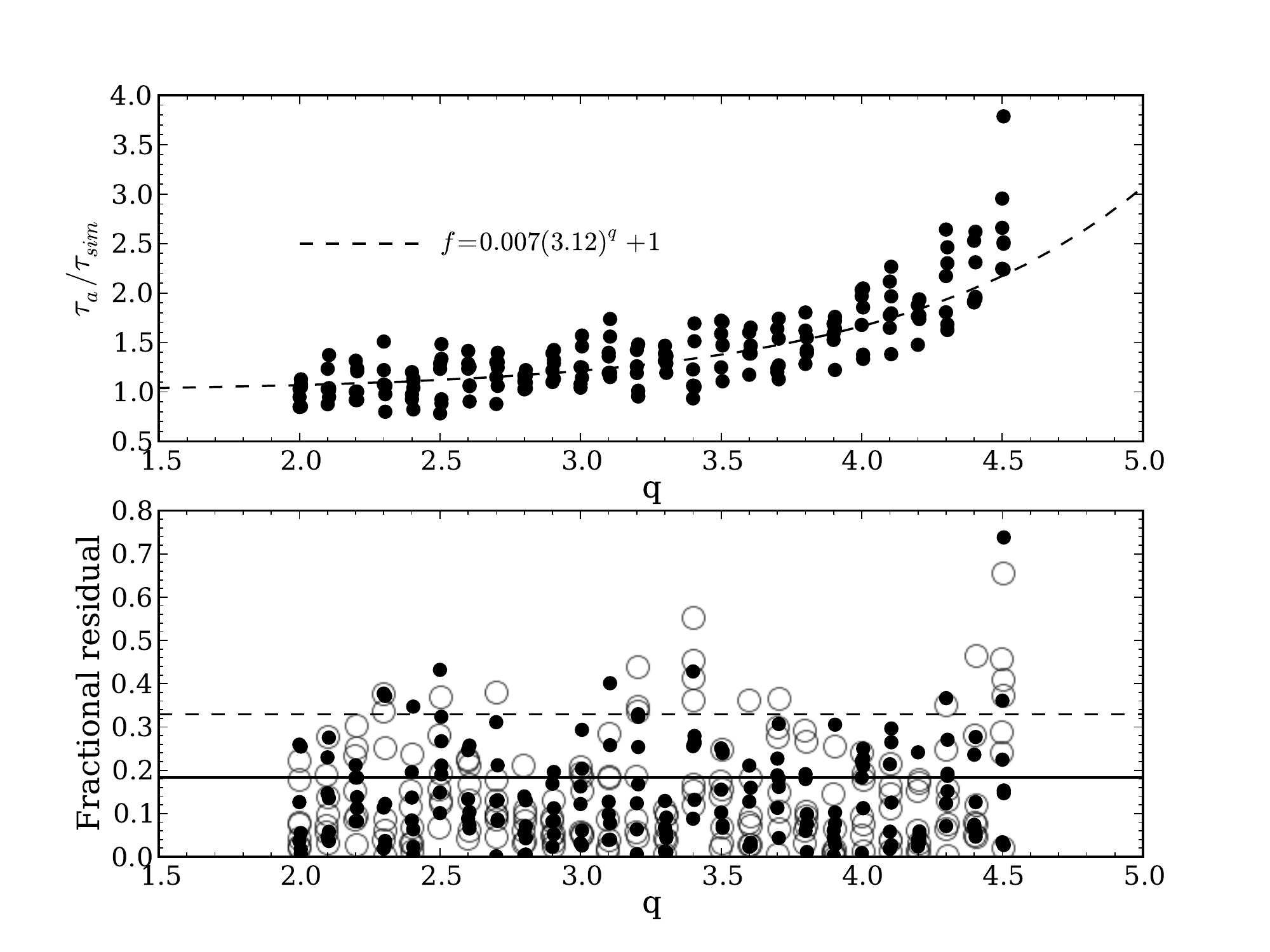}
\caption{Same as Figure \ref{tau_corr}, but for simulations with $V_i = 2$ km s$^{-1}$ and $\rho= 0.4$ g cm$^{-3}$. Note the somewhat poorer performance of the correction function $f$ at extremely steep slopes due to increased mass loss; however, the single power-law fit to the system lifetime vs. $q$ also performs poorly at these high slopes.}
\label{tau_corr_highv}
\end{centering}
\end{figure}

Figure \ref{tau_corr} illustrates $\tau_a / \tau_{sim}$ for all seven ultra-wide TNBs characterized in this work, and shows Eqn. \ref{fit_corr} for comparison. We stress that this correction has no rigorously physically-motivated form; other functional forms were explored, but in general did not improve the scatter significantly. No obvious trend in initial system separation, mass, or eccentricity were found when attempting to reduce the scatter in the correction results. We have also compared this correction to simulations with different impact velocity ($V_i = 0.5-2$ km s$^{-1}$), different bulk density ($\rho=0.4-2$ g cm$^{-3}$), and simulations of much more tightly bound binaries ($a_m /R_H \simeq 0.01$) and find that it generally holds in these regimes as well. The correction is found to break down at high size-distribution slopes ($q\gtrsim4$) in cases of simultaneously high impact velocity (2 km s$^{-1}$), low density (0.4 g cm$^{-3}$), and tightly bound binaries ($a_m /R_H \simeq 0.01-0.02$) because the amount of mass-loss suffered before these systems become unbound tends to be rather large; as such, the true system lifetimes tend to be somewhat shorter than predicted by the analytical correction in these extreme cases. Figure \ref{tau_corr_highv} illustrates the same quantities as Figure \ref{tau_corr} for the ultra-wide binaries in a set of simulations with $V_i = 2$ km s$^{-1}$ and $\rho=0.4$ g cm$^{-3}$; in general the same correcting function still performs better than the Eqn. \ref{tau0} power-law fit to $\tau$ vs. $q$, though the correction performs less well in this case than in the $V_i = 1$ km s$^{-1}$, $\rho=1$ g cm$^{-3}$ case.

Combining Eqns. \ref{tau_analytic} and \ref{fit_corr} allows for accurate estimate of nearly any TNB,

\begin{equation}\label{final_tau}
\tau_{corr} \simeq \left( (0.007 \times 3.12^{q} + 1) P_{i} N_0 R_0^{q-1} \left( R_s^{3-q} + R_p^{3-q} \right) \left( \frac{ 0.62 }{ V_i} \right)^{\frac{1-q}{3}} \left( \frac{G M_{sys}  }{a_m} \right)^{\frac{1-q}{6}} \right)^{-1}.
\end{equation}

For all further discussion, system lifetimes are estimated as $\tau = \tau_{corr}$.

\subsection{Small object population limits}

Given the current existence of a population of $n$ objects today with individual mean lifetimes $\tau_{i}$, the initial population implied by that population is 

\begin{equation}\label{sum}
n_{0}= \sum\limits_{i=0}^n e^\frac{t}{\tau_{i}}
\end{equation}

\noindent where $t$ is the time over which the binary population has been decaying, estimated as $4\times10^9$ years. Given a maximum initial population of binaries, we can then numerically solve Eqn. \ref{sum} given $\tau_{i}$ determined by Eqn. \ref{final_tau} and the parameters in Table 1. The parameter we choose to vary in Eqn. \ref{final_tau} in order to solve Eqn. \ref{sum} is $N_0$ which we take to be $N(R > 1 \mbox{ km})$ for a given power-law slope $q$. This results in an estimate of the maximum population of $R>1$ km objects in the classical belt allowed by the continued existence of our sample of ultra-wide binaries, given size distribution slope $q$ and an assumed initial population of binaries. This estimate also assumes that the impactor size distribution is in equilibrium, and is not evolving in time. Thus, we take these estimates to reflect the current collisional environment of the Kuiper Belt ignoring any early collisional evolution of the size distribution.

Given a current binary fraction $f$ and a primordial binary fraction $f_0$, the number of primordial binaries $n_0$ implied by a given number of extant binaries $n$ is

\begin{equation}\label{primordial_number}
n_0 = n \frac{f_0}{f},
\end{equation}

As our most conservative estimate, we assume that $\sim$100\% of the current Cold Classical objects started their lives as binaries. At present $\sim30$\% of Cold Classical objects exist as tight to moderately-wide binaries (Noll et al. 2008a), so we set the primordial ultra-wide binary fraction to be at most the remaining 70\%. The current fraction of ultra-wide binaries is estimated to be much lower, approximately 1.5\% (Lin et al. 2010). To be conservative, we assume that after disruption of a binary system, only one of its components remains as a Cold Classical object (the other being lost from the population). Using these binary fractions, the $n_0$ implied by our sample of seven ultra-wide binaries is

\begin{equation}
n_{0} =  7 \frac{0.7}{0.015} = 327 \mbox{ (Case 1).}
\end{equation}

However, such a high fraction of ultra-wide binaries is not physically motivated by any formation model. To set a more realistic upper limit of the impactor population, we consider the results of the binary formation simulations presented by Nesvorn\'{y} et al. (2010). These simulations modeled the formation of binaries through gravitational collapse, and Parker et al. (2011) found that while somewhat over-producing ultra-wide binaries compared to the extant sample, the orbital distribution produced was favorably similar to the observed distribution. The relative fraction of ultra-wide binaries (splitting at $a_m /R_H = 0.07$) produced by these simulations is approximately 20\% of all binaries formed. Therefore, even if the total binary fraction of the primordial Cold Classical Kuiper Belt was 100\%, the primordial ultra-wide binary fraction likely did not exceed 20\% given this formation scenario. Using this primordial binary fraction, the $n_0$ implied by our sample of seven ultra-wide binaries is

\begin{equation}
n_{0} =  7 \frac{0.2}{0.015} = 93  \mbox{ (Case 2).}
\end{equation}

\begin{figure}
\begin{centering}
\includegraphics[width=1\textwidth]{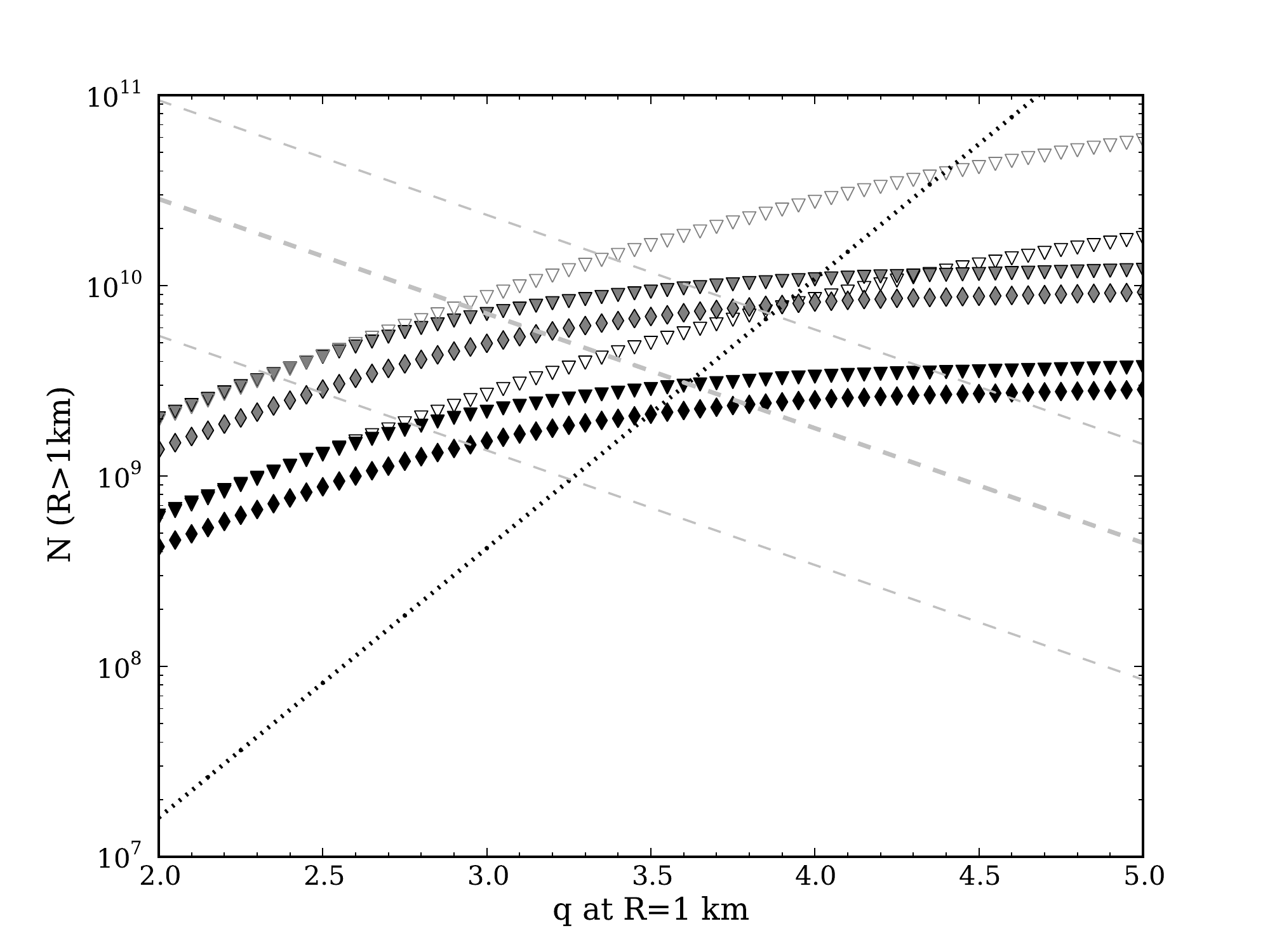}
\caption{Upper limits on number of objects larger than 1 km that pass through the Classical Kuiper Belt, assuming ultra-wide binaries today represent an eroded primordial population. Results assuming $P_{i}=1.3\times10^{-21}$ km$^{-2}$ yr$^{-1}$ (black points) and $P_{i}=4\times10^{-22}$ km$^{-2}$ yr$^{-1}$ (gray points). Triangles show upper limit for case 1 (70\% primordial ultra-wide binary fraction), while diamond points show upper limit for case 2 (20\% primordial ultra-wide binary fraction). Open triangles show limit when 2000 CF$_{105}$ is removed for case 2. Dotted line shows extrapolated population assuming albedo of $p=0.1$, determined as described in the text with Eqn. \ref{extrap_pop}. Dashed silver lines shows best-fit (heavy line) and 1-$\sigma$ limits (light lines) extrapolated from the estimate of the $R\gtrsim 250$ m population by occultations (Schlichting et al. 2009).}
\label{limit}
\end{centering}
\end{figure}

We set $n_0$ in Eqn. \ref{sum} first equal to 327 (case 1), then to 93 (case 2) and solve for $N(R > 1 \mbox{ km})$ as a function of $q$ given the collisional lifetimes from Eqn. \ref{final_tau}, assuming two values of $P_i$;  $1.3\times10^{-21}$ km$^{-2}$ yr$^{-1}$ suggested by Farinella et al. (2000) for collisions between Classical Kuiper Belt objects, and $4\times10^{-22}$ km$^{-2}$ yr$^{-1}$ as derived by Dell'Oro et al. (2001) for the same circumstances but assuming a different orbital distribution and using a different derivation technique. The results are illustrated in Figure \ref{limit}; under any of the size distribution slopes we consider, the population of 1 km radius objects in the Classical Kuiper Belt must be less than $\sim2\times10^{10}$ objects (assuming the smaller $P_i$ value) or less than $5\times10^9$ objects (assuming the larger $P_i$ value), with fewer objects being allowed for lower $q$.

For comparison, we extrapolate the measured large-object population to $R\simeq1$ km. The CFEPS L7 synthetic model of the Kuiper Belt\footnote{Available at \url{http://www.cfeps.net/L7Release.html}} contains $\sim45,500$ objects with $H_g<8.5$ in the Main Classical Kuiper Belt (hot, stirred, and kernel components). Extrapolating this population to a break magnitude of $H_g = 10$ with a luminosity function slope of $\alpha=0.76$ (eg., Fraser \& Kavelaars 2009) we find $\sim 618,000$ in this population larger than the break magnitude (translated with $p=0.1$ to a radius of 26 km). We then extrapolate this number to $R=1$ km using a size distribution with slope $q$ normalized at $R=26$ km,

\begin{equation}\label{extrap_pop}
N(R>1 \mbox{ km}) = 618,000 \times \left(\frac{ 1 \mbox{ km} }{ 26 \mbox{ km} } \right)^{1-q}.
\end{equation}

Current observations suggest that the Cold Classical Kuiper Belt size distribution breaks at radii of 20-30 km to a slope of approximately $q\simeq2$ (Bernstein et al. 2004, Fraser \& Kavelaars 2009, Fuentes et al. 2009). If this slope continues all the way down to radii of 1 km, the implied population of impactors would allow the survival of a relic ultra-wide binary population over the age of the Solar System. However, such a slope would be inconsistent with the putative detection of a single stellar occultation event by a $\sim250$ m TNO reported by Schlichting et al. (2009). The convergence of the $R>1$ km population estimates at slopes of $q\sim3.5$ from extrapolating the large-object population, collisional lifetimes of binaries, and stellar occultations, combined with the fact that a the total mass of a population with size distribution slope steeper than $q = 4$ is infinite, suggests that the small-object size distribution slope lies between $3.2 \leq q < 3.8$, and $N(R>1$ km) is a few billion. Using the less conservative $P_i = 1.3\times10^{-21} km^{-1} s^{-1}$, the maximum viable slope becomes $q \lesssim 3.6$ for case 1 and $q \lesssim 3.5$ for case 2.

At the steep-slope end, the current existence of the binary 2000 CF$_{105}$ sets the strongest upper limit on the impactor population, as its low mass and wide separation make it extremely easy to disrupt by the numerous small impactors in this regime. Figure \ref{limit} includes the upper limit on the impactor population with 2000 CF$_{105}$ removed, and while at low $q$ there is no change, at high $q$ the upper limit becomes much less constraining. Small binaries like 2000 CF$_{105}$ are at present the least complete sample, as they suffer the strongest flux bias --- only those with very high albedos are detected in current surveys, and 2000 CF$_{105}$ likely represents the first of a large population. Determining the prevalence of small 2000 CF$_{105}$-like binaries in the current Kuiper Belt should therefore be a critical goal of future large-scale surveys.

Note that these estimates remain fairly conservative; even in case 2 (primordial ultra-wide binary fraction of 20\%) we assume that the total primordial binary fraction was 100\% and no intense period of collisional grinding occurred (that is, the impactor population has not decayed significantly over the age of the solar system). Such an epoch would be extremely destructive to the primordial binary population (Petit \& Mousis 2004, Nesvorn\'{y} et al. 2011). Additionally, we assumed in each case that when a binary was disrupted, only one of its two components survived on as a solitary TNO; in reality, a large fraction of both the disrupted primaries and secondaries would survive as independent TNOs in this population. If we assume both components survive, $n_0$ derived from Eqn. \ref{primordial_number} becomes

\begin{equation}
n_0 = n \frac{1 + f^{-1} }{1 + f_0^{-1}},
\end{equation}

\noindent which produces $n_0 = 195$ for case 1 and $n_0 = 79$ for case 2. 

Also clear from Figure \ref{limit} is that these estimates are quite sensitive to the adopted value of the intrinsic collisional probability. In fact, the upper limit on the small-object population is inversely proportional to the adopted value, as $\tau \propto (P_{i} N(R>1\mbox{ km}) )^{-1}$, and therefore for a fixed lifetime, $N(R>1\mbox{ km}) \propto P_{i}^{-1}$. Our uncertainties on the upper limits on the small-object population are driven largely by the uncertainty in $P_{i}$ for today's Kuiper Belt; the difference between the two values adopted in this work imparts a factor of 3.25 variation between the their respective estimates of the small-object population. Fortunately, when newer and more accurate values are available the population estimates we present can be revised by simply scaling them by the ratio of our adopted values of $P_i$ to the newer value.

\subsection{ Does orientation play a role in survival time?}\label{inc_survival}

The asymmetry in the maximum stable tidal radius between prograde and retrograde orbits allows wider orbits to exist stably for retrograde orbits, but we have found that this additional stable phase space does not significantly enhance the lifetime of retrograde binaries subjected to collisions drawn from realistic impactor size distributions.

By determining the change in velocity required to lift an initial orbit to an arbitrary final semi-major axis, and using this $\Delta v$ to determine the impactor size required to affect this change (under the assumption of a perfectly inelastic collision with the binary's secondary), the ratio of the population of impactors capable of lifting a given binary to beyond its prograde tidal limit to the population capable of lifting the same binary to beyond its retrograde tidal limit is approximately

\begin{equation}\label{asym}
\frac{N(>R_{p}) }{ N(>R_{r})} \simeq \left( \frac{ 1- a_0 / R_H } { 1 - 2 a_0 / R_H }\right)^{(q-1)/6},
\end{equation}

\noindent where $a_0$ is the initial semi-major axis, and assuming that the stable tidal limit for retrograde orbits is one classical Hill radius $R_H$, while the stable limit for prograde orbits is one-half of the classical Hill radius. Since the lifetime of the binaries is inversely proportional to the population of impactors that are capable of disrupting them, this ratio represents the ratio of lifetimes of retrograde and prograde binaries. Comparing this relationship to the known ultra-wide binaries ($a_0/R_H \sim$ 0.08---0.25), we see that even with extremely steep impactor size distributions this ratio is close to unity, and thus would expect very little asymmetry between the expected mean lifetimes of the prograde and retrograde binaries.

To verify that survival time is generally independent of inclination, we performed a test of the effect of the tidal stability asymmetry by re-running our numerical simulations and approximating the tidal stability limits as 

\[ R'_H = \left\{ \begin{array}{ll}
         0.5 R_H & \mbox{if $i_m \leq 90^\circ$};\\
         R_H & \mbox{if $i_m > 90^\circ$}.\end{array} \right. \]

We selected the initial inclination for each binary system from a uniform distribution between $0^\circ < i_0 < 180^\circ$, and determined whether each binary's mutual apocenter remain below the our approximate of the stable limit for its current inclination after every collision. The resulting mean lifetimes (for values of $q$ ranging from 2---4.5) showed no discernible variation with initial system inclination.

The lack of strong variation indicates that if the primordial populations of prograde and retrograde populations were equal, that equality should persist to the present day. If there was any primordial asymmetry, however, evolution of the mutual inclination may cause some system's orientation to flip, thereby causing the prograde-to-retrograde ratio to change over time. We explore this possibility in the following section.

\section{Evolution of orbital parameters}

\subsection{Evolution of the inclination distribution}

In our collisional simulations, we track the initial mutual inclination and the final inclination the system reaches before being disrupted. A significant unknown in the current understanding of how to interpret the orbital distribution of TNBs is how inclination evolves over time, as binary systems are subjected to various perturbations; in the case of collisional perturbations, we can determine the effect directly. In the following experiments, we considered impactor populations with a fixed size distribution slope of $q=3$. We adopt this slope because it is still allowed by the arguments of the previous section when considering extrapolation of the $R\sim1$ km population from the measured population of large objects, and because steeper slopes allow more chance of stochastic evolution of orbital evolution through multiple small impacts.

\begin{figure}
\begin{centering}
\includegraphics[width=0.5\textwidth]{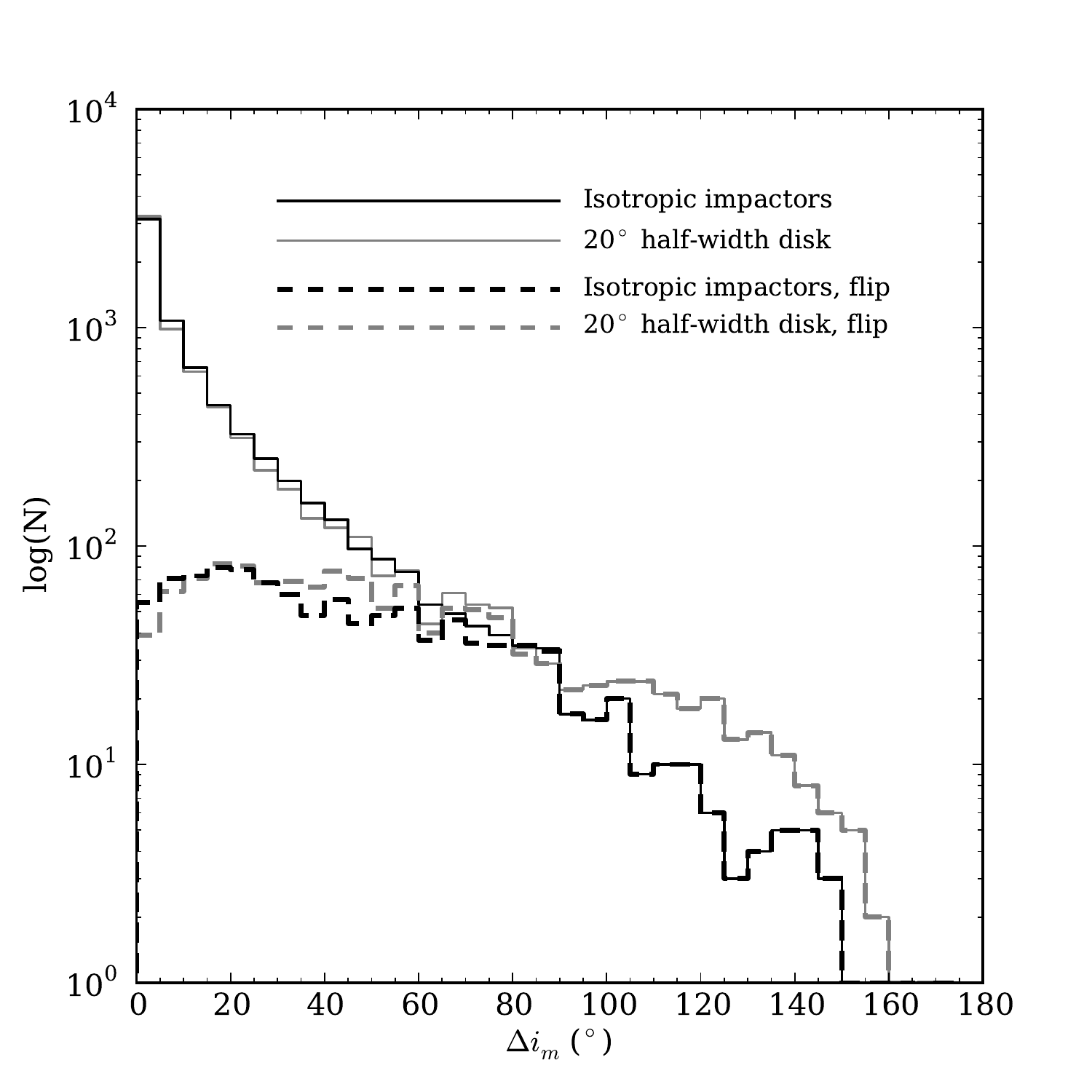}
\caption{Histogram of the change in inclination before disruption of binary systems. Initial inclinations were drawn from a uniform distribution, and impactors either struck from random orientations or within a disk of half-width $20^{\circ}$.  Dashed histograms illustrate the just those system's which had their orientation reversed from prograde to retrograde (or vice versa). 15---18\% of systems are reoriented, with disk-like impactor geometry more efficient.}
\label{delta_inc_u}
\end{centering}
\end{figure}

Figure \ref{delta_inc_u} shows the difference between the initial and final mutual inclinations for 100 realizations of each binary system, where the initial inclinations were drawn from a uniform distribution ($p(i) \propto \sin(i)$) and impactors either struck uniformly from all directions or were drawn from a longitudinally-uniform disk with half-width of $20^\circ$. Final inclination was determined to be the last inclination of the system prior to the final impact that disrupted it. Many systems have little to no change in inclination before disruption; however, significant change did occur for some systems, and $\sim$15\%---18\% of the systems had their orientation flipped from prograde to retrograde or vice-versa, with the disk-like geometry more efficient at reversing orientations. The final inclination distribution was found to be indistinguishable from the initial, uniform distribution in both cases; in general, random perturbations will tend to make a non-uniform distribution \textit{more} uniform, and not vice-versa. 

\begin{figure}
\begin{centering}
\includegraphics[width=0.5\textwidth]{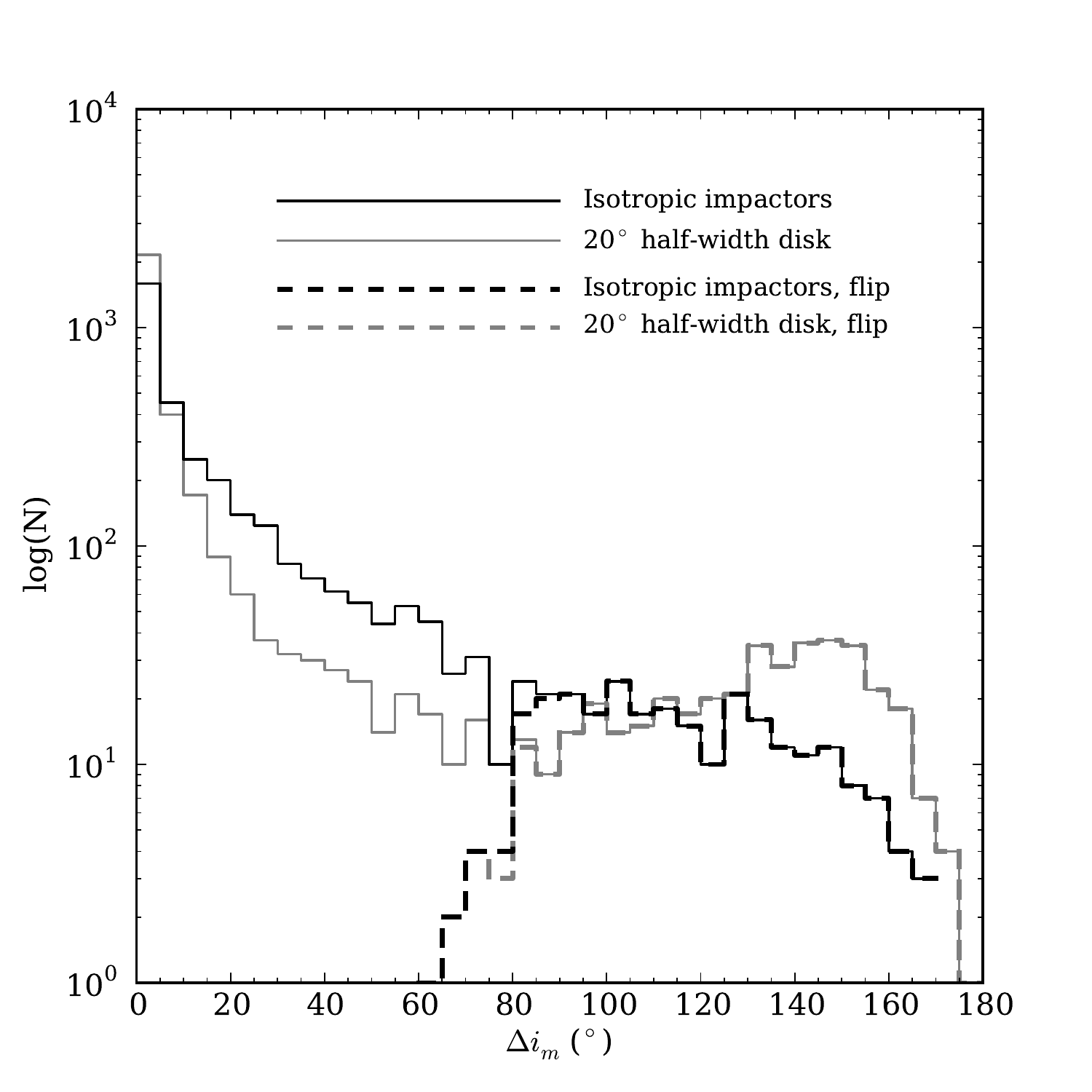}
\caption{Same as Figure \ref{delta_inc_u}, but with initial inclinations drawn from $p(i) \propto \sin(i)e^{-\frac{1}{2}\left( \frac{i}{10^{\circ}} \right)^2}$. Fewer systems are reoriented, at between 7-11\%. }
\label{delta_inc}
\end{centering}
\end{figure}

The inclination distribution of TNBs has been shown to be indicative of formation mechanics (eg., Schlichting \& Sari 2008), and it has been measured by recent surveys. The ultra-wide binary inclination distribution is currently inconsistent with being drawn from a uniform distribution (Parker et al. 2011). It lacks any high inclination systems ($55^\circ \lesssim i_m \lesssim 125^\circ$), and has a large number of systems at very low mutual inclination. This preference for pole-aligned mutual orbits is suggestive of formation in a dynamically cold disk (eg., Noll et al. 2008b). In order to determine the evolution of a primordially cold inclination distribution, we repeated our collisional simulations with an initial inclination distribution given by a sine times a gaussian, centered at $i=0^\circ$ with a width of $\sigma=10^\circ$,

\begin{equation}
p(i) \propto \sin(i) e^{-\frac{1}{2}(\frac{i}{10^\circ})^2},
\end{equation}

\noindent and determined the final inclination distribution given impactors striking uniformly from all directions or drawn from a longitudinally-uniform disk with half-width of $20^\circ$. The results are illustrated in Figure \ref{delta_inc}. Both final inclination distributions are strikingly similar to the present inclination distribution of the ultra-wide binaries. In the case of randomly-oriented impact trajectories, more inclination evolution occurred regardless of initial mutual inclination; however, in the case of disk-like geometry for the impactors, systems with low initial inclination tended to suffer less inclination evolution than the rarer systems with higher initial mutual inclination. A smaller fraction of systems changed orientation than in the case of uniform initial inclinations, due to the average system having to suffer significantly larger excursions in inclination in order to change orientation. Randomly-oriented impacts caused roughly 7\% of systems to reorient, while impact trajectories drawn from a disk-like distribution caused roughly 11\% to reorient.

\begin{figure}
\begin{centering}
\includegraphics[width=0.5\textwidth]{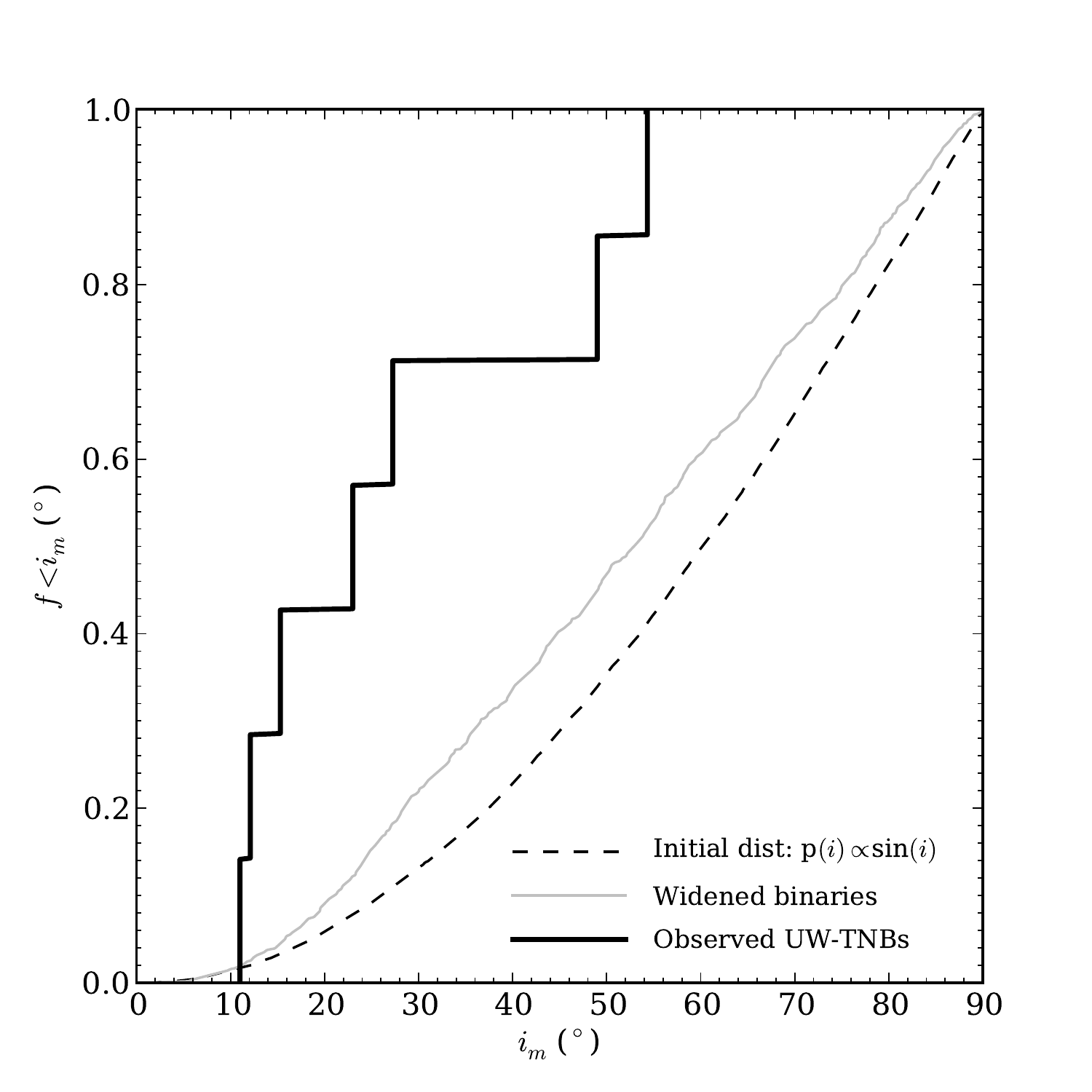}
\caption{Comparison of initial and final inclination distributions after collisional evolution (final inclination taken just prior to binary disruption). Initial binary taken to have $a_m /R_H = 0.02$, and only those systems which are widened to $a_m /R_H>0.07$ before disruption are considered. Initial inclinations drawn from $p(i) \propto \sin(i)$ (dashed line), and final inclinations of widened systems are shown for disk-like impactor geometry (gray line). Distribution of final inclinations for random impact trajectories are identical to initial distribution. Current ultra-wide TNB inclinations shown for comparison (heavy black line); that their inclinations are drawn from distribution of widened binary sample is ruled out at $>95$\% confidence. }
\label{tight_inc}
\end{centering}
\end{figure}

We also investigated the possibility that with a disk-like impactor population, binaries which suffer collisions which widen their orbits will preferentially have their mutual inclinations decreased; this can be understood by thinking of ``stretching'' an initially inclined orbit in the plane of the impacts. We took synthetic binaries with initial $a_m /R_H \sim 0.02$ and subjected them to collisions with the same impactor populations as our ultra-wide binary experiments, but only with impact trajectories drawn from the disk-like distribution. We drew their initial inclinations from a uniform distribution, then considered the final inclinations of only those systems which had become widened (prior to disruption) to $a_m /R_H \gtrsim 0.07$, comparable to the minimum separation of the ultra-wide binaries considered here. Since these tighter systems take longer to disrupt than the ultra-wide systems, they were actually given a longer time over which to have their inclination modified than the ultra-wide binaries in our simulations. Figure \ref{tight_inc} illustrates that while the inclination distribution does evolve away from the uniform distribution (unlike the behavior that would be expected from randomly-oriented impact trajectories), the inclination distribution does not change significantly enough to make this mechanism feasible for explaining the current inclinations of the ultra-wide binaries. The KS statistic rules out that the ultra-wide binary inclination distribution was drawn from this collisionally-modified uniform distribution at greater than 95\% confidence.

We conclude that if the ultra-wide binaries represent a highly collisionally-evolved population, then they must have had a much colder primordial inclination distribution. Additionally, they cannot have evolved from tighter binaries, because the current inclination distribution of tighter systems are close to uniform (Grundy et al. 2011, Parker et al. 2011) and collisions cannot produce a widened population with as cold an inclination distribution as is observed for the ultra-wide binaries when starting with an initially uniform inclination distribution.

\subsection{Evolution of separation and eccentricity}

\begin{figure}
\begin{centering}
\includegraphics[width=1\textwidth]{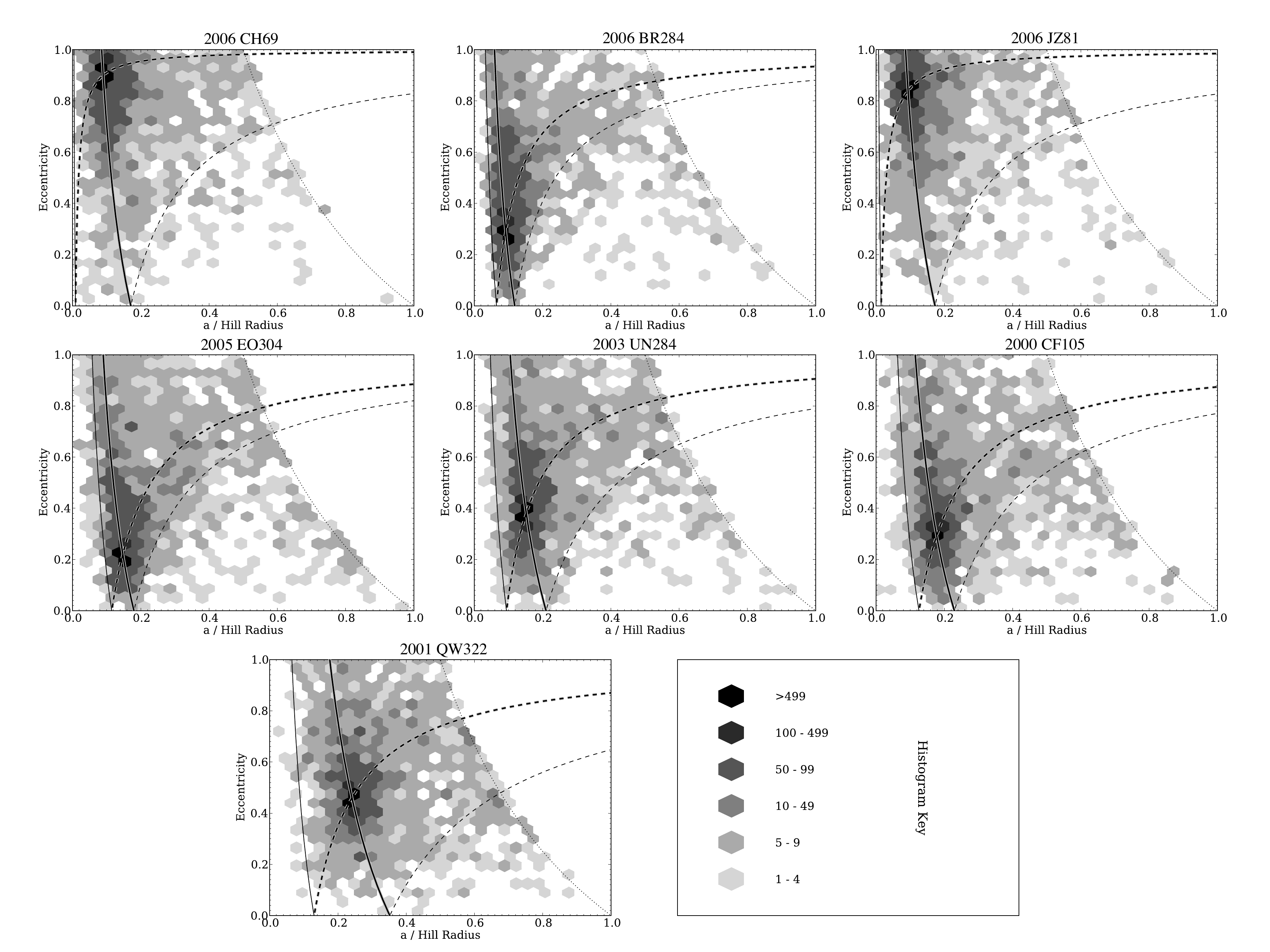}
\caption{2D histogram of last orbit before collisional disruption for each ultra-wide binary characterized, sorted in order of increasing initial $a_m /R_H$. These panels show outcomes for simulations with $q=3$. Heavy solid line is constant apocenter ($e' = (a_0/a')(1+e_0)-1$). Heavy dashed line is constant pericenter ($e' = 1-(a_0/a')(1-e_0)$). These lines cross at the systems' current $a_m / R_H$ and $e$. Light dashed line is where initial apocenter is final pericenter ($e' = 1-(a_0/a')(1+e_0)$), and light solid line is where initial pericenter is final apocenter ($e' = (a_0/a')(1-e_0)-1$). These lines mark the region within which a single non-unbinding collision can drive a binary. Note that binaries prefer to evolve along line of constant apocenter. Dotted line marks apocenter larger than the Hill radius, our criteria for disruption.}
\label{aRH_evol}
\end{centering}
\end{figure}

In addition to tracking the inclination of the ultra-wide binaries during our collisional simulations, we also track their semi-major axis and eccentricity. Figure \ref{aRH_evol} illustrates the distribution of $a_m /R_H$ and $e$ just prior to collisional disruption for each binary system simulated, again with size-distribution fixed with $q=3$. Since collisions preferentially occur while the system is at mutual apocenter, orbital evolution prefers to occur along the line of constant apocenter: $e' = (a_0/a')(1+e_0) -1$. The most severe increase in semi-major axis that can occur in a single non-unbinding collision is the condition where the original apocenter becomes the system's new pericenter: $e' = 1 - (a_0/a')(1+e_0)$. Evolution to wider separation requires at least two significant collisions, and we find that it is uncommon for a system to be subjected to two such collisions while remaining bound. Between $90-95$\% of all final orbits have final pericenters lower than their initial apocenter, and the most common behavior is to have next-to-no significant evolution prior to the collision which disrupts the system.

This conservation rule of $q' \leq Q_0$ can be used to further argue against the possibility that the current ultra-wide binaries represent a collisionaly-widened tail of the tight binary population. If this were the case, not only would we expect a randomized inclination distribution, but we would also expect relatively high eccentricities to be the rule among the ultra-wide binaries. For an initially circular binary with initial $a_0/R_H = 0.02$, this conservation rule would state that if the system was widened to $a'/R_H = 0.1$ its eccentricity would usually exceed $e' \gtrsim 0.8$. As only two of the seven binaries in our sample exceed this eccentricity, it is unlikely that the ultra-wide binaries are the outcomes of this kind of evolution. This is in line with the results of Nesvorn\'{y} et al. (2011) who also found that it is unlikely for primordially tight binaries with primary radii of $\sim50-100$ km to be collisionally widened to the presently observed separations of the ultra-wide binaries.

In fact, because of the larger available phase-space at high eccentricity, evolution along constant apocenter tends to increase the system's eccentricity for systems with low to moderate initial eccentricity while \textit{decreasing} separation. Thus, systems like 2006 CH$_{69}$ with extremely high mutual eccentricities may represent the outcomes of collisional modification of initially wider and less eccentric systems. This is an attractive prospect, as the mutual pericenter passages of 2006 CH$_{69}$ during the high-eccentricity phases of its Kozai cycles ($q\simeq31R_P$, Parker et al. 2011) may be close enough to cause significant orbital shrinking and circularization over the age of the solar system; if instead its present eccentricity is the outcome of a relatively recent collision, then it need not have maintained such close pericenter passages over such a long period.

\section{Discussion}

\subsection{The curious case of 2000 CF$_{105}$}

As discussed in \S3.2, the binary 2000 CF$_{105}$ is the most susceptible to collisional disruption under steep size distributions, due to its wide separation and very small component sizes. Its current existence places the largest constraint on the population of impactors for these high slopes, but since at present it is only one binary the level of confidence one should have in this constraint is not immediately clear. For example, there is a non-zero probability that it represents a system which was primordially tightly-bound which has been anomalously widened, and it has not somehow survived over the age of the solar system in its current configuration.

However, a number of aspects of 2000 CF$_{105}$ lead us to conclude that this case is very unlikely. First is its small size; it fell within the flux limits of current surveys only because of its extremely high albedo ($p \simeq 0.3$). Parker et al. (2011) argue that somewhat lower albedos appear to be more common in the Kuiper Belt, and if this is the case then there is likely a large population of low-albedo 2000 CF$_{105}$-like binaries lurking unseen beneath the flux limits of current surveys. In other words, because of our current relative insensitivity to binaries of its size, the detection of 2000 CF$_{105}$ suggests that binaries of similar size are intrinsically common.

Additionally, if 2000 CF$_{105}$ were a collisionally-evolved tight binary, we would expect it to have both a high eccentricity and a mutual inclination drawn from the same uniform distribution as is observed for the tighter binaries (Parker et al. 2011, Grundy et al. 2011). However, its eccentricity is one of the lowest in our sample at $e=0.29$, and it has the second-most aligned mutual orbit pole of any TNB known with a mutual inclination of 167.9$^{\circ}$. Randomly drawing such an inclination from a uniform distribution is extremely unlikely ($p(|i_m|<13^\circ)\simeq 0.026$). 

Together, the fact that 2000 CF$_{105}$ is likely the harbinger of many more small-radius binaries and that its current mutual orbit  appears inconsistent with being generated by widening a tightly-bound binary through collisions, we conclude that the assumption that it has existed as an ultra-wide TNB in a configuration relatively similar to its present state over the age of the solar system is merited. By extension, we have confidence that its continued existence is a valid constraint on the small-object TNO population for steep size distributions.

\subsection{Implications for formation mechanisms}

In Parker et al. (2011), the ultra-wide binary inclination distribution was found to be inconsistent with a uniform distribution due to its preference for inclinations aligned with outer orbit poles. Such an inclination distribution was found to be suggestive of formation in a dynamically-cold disk. However, because the orientations of the ultra-wide binaries were found to be consistent with no preference for prograde or retrograde, and they ruled out an extreme preference for retrograde orientations predicted by Schlichting \& Sari (2008) for formation by the $L_2 s$ mechanism in a very cold disk, it was concluded that at the time of binary formation the velocity dispersion of the disk must have been approximately the Hill velocity. 

In this work, we have found that if the ultra-wide binary inclination distribution is in fact non-uniform, it must have been \textit{less} uniform in the past with stronger preference for low inclinations. However, we have also found that if subjected to a maximally-erosive disk over the age of the solar system, a non-negligible fraction ($7-11\%$) of these systems systems could be reoriented from prograde to retrograde and vice-versa. If the primordial prograde-to-retrograde ratio was $\sim 0.03$ as predicted by Schlichting \& Sari (2008) for formation in a very dynamically-cold disk, then reorienting 10\% of the binaries over the age of the solar system would result in a current prograde to retrograde ratio of $\sim 0.14$. Even after this reorientation, however, the probability of randomly sampling 4 prograde and 3 retrograde systems from a distribution with a prograde-to-retrograde ratio of 0.14 is less than 1\%. Thus it seems that collisions cannot be invoked to reorient a sufficient number of binaries to make the $L_2 s$ mechanism a viable explanation for the majority of ultra-wide binary orbits.

Parker et al. (2011) also explored the possibility that these ultra-wide binaries formed through the gravitational collapse mechanism posited by Nesvorn\'{y} et al. (2010), and found generally encouraging results. However, it was found that when correcting for observational completeness, the Nesvorn\'{y} et al. (2010) results somewhat over-produce ultra-wide binary systems (roughly by a factor of four) with respect to the currently observed separations. Given collisional decay of both populations, we can estimate how the ratio of wide to tight binaries would evolve over the age of the solar system:

\begin{equation}\label{wide2tight}
\frac{N'_{wide} } {N'_{tight}} = e^{t  \left(\frac{1}{\tau_{tight}} -\frac{1}{\tau_{wide}} \right)}\frac{N_{wide,0} } {N_{tight,0}} = \left( \frac{f_{tight,0} } {f'_{tight}} \right)^{1 - \frac{\tau_{tight}}{\tau_{wide}} }\frac{N_{wide,0} } {N_{tight,0}},
\end{equation}

\noindent where $N_{x,0}$ represents the primordial number of a given population $x$, while $N'_{x}$ represents the current number of that same population $x$ after collisional decay, and $f'_{tight}$ and $f_{tight,0}$ represent the current and primordial fraction of tight binaries. Using Eqn. \ref{tau_ratio_aRH} to estimate $\frac{\tau_{tight}}{\tau_{wide}}$ for binaries with the same mass but a ratio of $( a_{tight} /R_H)/( a_{wide}/R_H) = 0.02/0.1 = 0.2$ for a size-distribution slope of $q=3$ gives a ratio of lifetimes of roughly 1.7. Comparing an even more tightly bound binary to a more widely separated one, $( a_{tight} /R_H)/( a_{wide}/R_H) = 0.01/0.2 = 0.05$, we find an even larger lifetime ratio of $\sim2.7$.

Substituting $\frac{\tau_{tight}}{\tau_{wide}}=1.7$ into Eqn. \ref{wide2tight} along with a primordial tight binary fraction of 80\% (based on the sub-sample of results of Nesvorn\'{y} et al. 2010 considered by Parker et al. 2011, and assuming a total primordial binary fraction of 100\%) and a current binary fraction of $\sim$29\% (eg., Noll et al. 2008a), we find that the fraction of wide to tight binaries could have been scaled down by roughly a factor of 2.6 over the age of the solar system. Using the larger lifetime ratio of $\frac{\tau_{tight}}{\tau_{wide}}=2.7$, we find that the fraction of wide to tight binaries could be reduced by over a factor of 10 over the age of the solar system --- however, this case would imply a current ultra-wide binary fraction of roughly half a percent, less than is observed. 

Somewhere between these two cases, erosion is sufficient to account for the roughly factor of four discrepancy between the predictions of Nesvorn\'{y} et al. (2010) and the current ratio of wide to tight binaries. A primordial tight binary fraction of 80\% and wide binary fraction of 20\%, scaled down to a present-day binary fraction of 29\% with a factor of four decrease in relative fraction of wide binaries to tight binaries would imply a present day wide binary fraction of roughly 1.4\%, slightly less than is observed. However, even this qualitative agreement is very encouraging given the somewhat preliminary nature of the simulations of Nesvorn\'{y} et al. (2010) and the simplicity of the analysis presented here.

\subsection{Rapid collisional grinding vs. slow erosion}\label{deplete}

Nesvorn\'{y} et al. (2011) showed that during a period of intense collisional grinding (motivated by the need to remove the excess mass required by hierarchical accretion models of planetesimal formation), a trend of decreasing binary fraction with decreasing radius would be imprinted on the surviving population in the radius range of the binaries considered in this work. Such behavior can be easily understood by considering the ratio of lifetimes expressed in Eqn. \ref{tau_ratio_aRH} and noting that for any impactor size distribution with slope steeper than $q=2.5$ system lifetime decreases with decreasing system mass. 

We can estimate the implied trend in binary fraction analytically. Given Eqn. \ref{tau_ratio_aRH}, the binary fraction with radius will be the following exponential:

\begin{equation}\label{binary_frac}
f(R) = 2^{-\left(\frac{R_{50}}{R} \right)^{(4q-10)/3}},
\end{equation}

\noindent where $R_{50}$ is the radius of a binary whose population will be reduced by 50\% due to collisional grinding after the elapsed time considered, given by solving Eqn. \ref{final_tau} for $R$ given $\tau = t / \ln(2)$ with $t$ being the elapsed time. To easily compare with the results of Nesvorn\'{y} et al. (2011), we convert Eqn. \ref{final_tau} into terms of $R_{50}$ and $a_m / R_H$ (assuming a binary with equal-mass components on a circular orbit about the Sun) and solve:

\begin{equation}
R_{50} = \left[ \frac{2 t (0.007 \times 3.12^{q} + 1) P_{i} N_0 R_0^{q-1} } {\ln(2)} \left( \frac{ 0.62 }{ V_i} \right)^{\frac{1-q}{3}} \left(  \frac{a_m}{R_H}  \frac {a_{\mbox{out}} }  {G ( \frac{ (8 \pi \rho )^2}{3} M_{\odot} )^{\frac{1}{3}}  } \right)^{\frac{q-1}{6}}    \right]^{\frac{3}{4q-10}}
\end{equation}

This simple analysis ignores breaks in the size distribution, but these can be included trivially by treating Eqn. \ref{binary_frac} in a piecewise manner:

\begin{equation}
f(R) = \left\{ \begin{array}{ll} 
f_1(R) = 2^{-\left(  \frac{R_{50}}{R}     \right)      ^{(4q_1-10)/3}} 									&: R > R_{c} \\
f_2(R) = \left(      f_1(R_{c}) \right)   ^{       \left(      \frac{R_{c}}{R}        \right)         ^{(4q_2-10)/3}} 		&: R \leq R_{c} \end{array}  \right. ,
\end{equation}

\noindent where $R_{c}$ is the primary radius of a binary who can be disrupted by impact with an object with radius $R_b$ which is the location of the break in the impactor size distribution between slopes $q_1$ (large object slope) and $q_2$ (small object slope). $R_{c}$ can be derived from Eqn. \ref{R_est}, and here we convert it to terms of $a_m / R_H$:

\begin{equation}
R_{c} = \left(  R_{b} \left(  \frac{V_i}{0.62} \right)^\frac{1}{3}  \left( \frac{a_m}{R_H} \frac  {a_{\mbox{out}} } { G ( \frac{ (8 \pi \rho)^2} {3}  M_{\odot} )^{\frac{1}{3}  }  }   \right)^{\frac{1}{6}}  \right)^{\frac{3}{4}},
\end{equation}

\noindent which, for a binary at 45 AU, with $V_i = 1$ km s$^{-1}$ and $\rho = 1 $ gram cm$^{-2}$, reduces to approximately

\begin{equation}
R_{c} \simeq 20.8 \times \left(  \frac{ R_{b} } { 1 \mbox{ km} } \right)^{\frac{3}{4}} \left( a_m / R_H \right)^{\frac{1 }{ 8 } }.
\end{equation}

\begin{figure}
\begin{centering}
\includegraphics[width=0.5\textwidth]{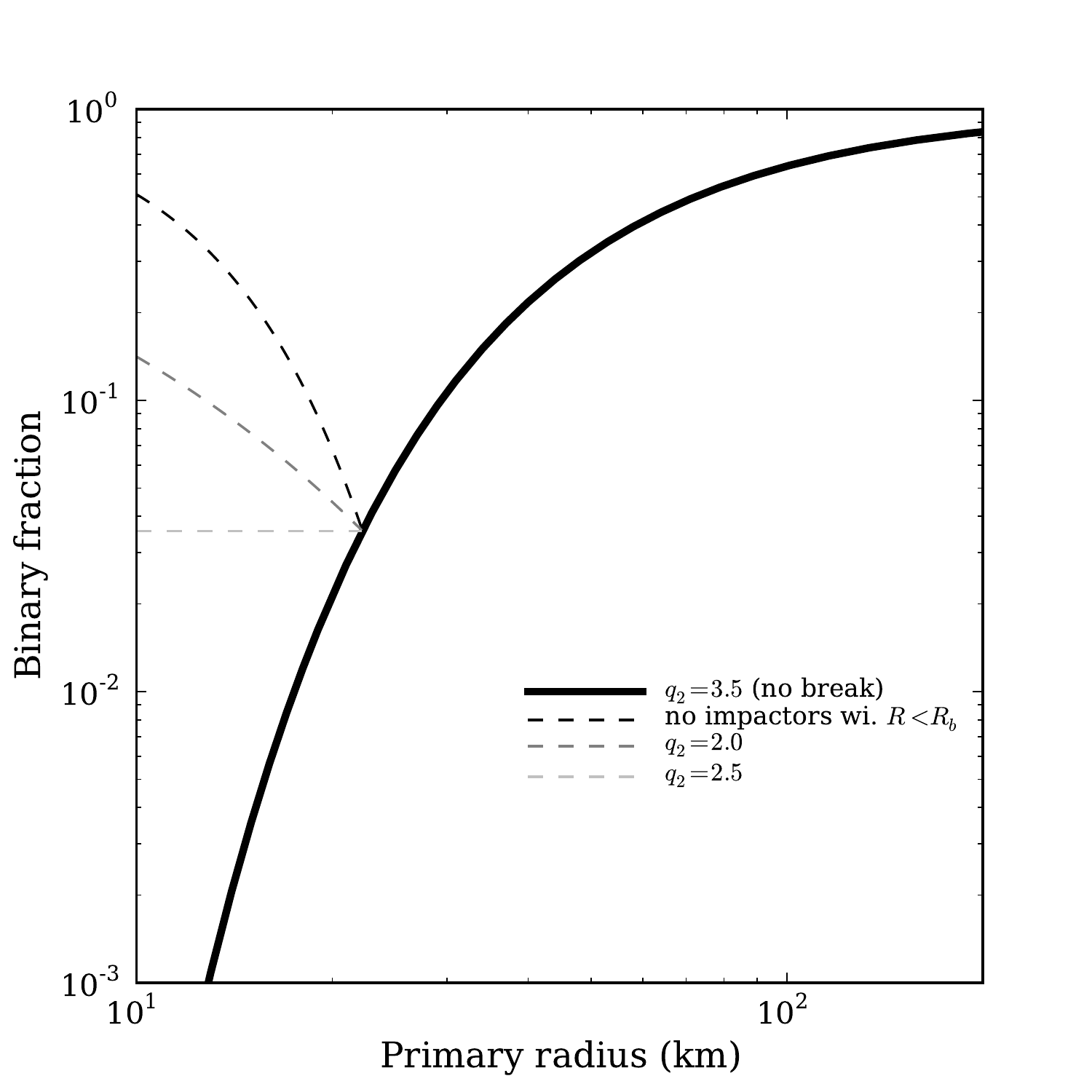}
\caption[Binary fraction with radius prediction]{Analytical estimate of the trend of binary fraction with primary radius, given an impactor population with size-distribution slope $q_1 = 3.5$ valid for objects larger than $R_b = 2$ km, while the slope for smaller objects is allowed to vary. This initial impactor size distribution (which has no break) adopts the following parameters: $N(R > 1$ km) $ = 5\times10^9$, $V_i = 1$ km s$^{-1}$, $P_{i} = 4\times10^{-22}$ km$^{-2}$ yr$^{-1}$, and assumes that the binaries all have separations $a_m/R_H = 0.1$ and are equal-mass systems. Elapsed time is taken to be $4\times 10^9$ years over which collisions have occurred. Heavy solid line shows trend with no break to shallower slope, while dashed lines show several cases with $q \leq 2.5$ where binary fraction will increase with decreasing radius.}
\label{binary_frac_trend}
\end{centering}
\end{figure}

As an illustrative example, we consider the expected binary fraction with radius given an impactor population with $N(R>1$ km)$ = 5\times 10^{9}$ and $q=3.5$ after $4\times10^9$ years have elapsed. We adopt $P_i = 4\times10^{-22}$ yr$^{-1}$ km$^{-2}$. Given these parameters and considering a binary with $a_m / R_H = 0.1$, we find $R_{50} \simeq  72.3$ km --- that is, over the age of the solar system, this impactor population would destroy 50\% of binaries with primary radius 72.3 km and initial $a_m / R_H = 0.1$. The resulting predicted binary fraction trend with radius is illustrated in Figure \ref{binary_frac_trend}.

Also illustrated in Figure \ref{binary_frac_trend} are the trends if there is a break arbitrarily added to the impactor size distribution at $R_{b} = 2$ km. As a limiting example, the trivial case where there are no impactors with radius less than 2 km is shown, as well as a break to a shallower size distribution with $q_2 = 2.5$ or $q_2 = 2.0$ below the break. With the break radius at $R_{b} = 2$ km, the primary radius where we would expect the binary fraction trend to change is $R_{c} = 26.2$ km. For the case of $q_2 = 2.5$, the binary fraction remains fixed for all radii smaller than $R_{c}$, while for the other cases the binary fraction climbs again for smaller radii.

Note that the resulting trends in radius, while not as strong, are very similar to the trends predicted by the numerical simulations presented in Nesvorn\'{y} et al. (2011) when considering a short period of intense collisional grinding. Though no strong evidence currently exists for a trend in binary fraction with radius, if such a trend is identified in the future further work will be required to disentangle its origin from one of two possibilities. Either such a trend could be produced through a period of strong collisional grinding in the early solar system, or it could be the result of slow collisional erosion over the age of the solar system (given a roughly steady-state impactor population with size distribution slope steeper than $q=2.5$). This complication lies atop the underlying issue that at present there is currently little theoretical constraint on what trends of binary fraction with radius might be produced by the binary formation mechanism directly, without any subsequent modification by collisional erosion.

One of the aspects of the Nesvorn\'{y} et al. (2011) simulations can also be treated analytically here; size-independent dynamical depletion of the impactor population. This is accomplished by noting that the impactor population normalization constant $N_0$ is in fact the time-average value, which reduces to the current value when considering an impactor population which is fixed in time. Thus, if the favored impactor population has been dynamically depleted by a factor $d$ over the age of the solar system, then the limits on $N_0$ made by the preceding arguments can be translated into the \textit{current} population limits by $N_{0(\mbox{current})} = N_0/d$. This estimate is valid only for size-independent depletion; the effects become much less trivial if the size distribution shape changes with time.

\subsection{Second-order effects: mutual tides and the Kozai effect}

This work generally ignored the effect of mutual tides (only treating them to the extent that mutual pericenter approaches within the Roche limit caused a merger), and did not include the effect of Kozai oscillations (Kozai 1962, Fabrycky \& Tremaine 2007, Perets \& Naoz 2009) on the evolution of the binary orbits. Because we found that the correction from the analytical estimate of collisional lifetimes did not vary systematically with system eccentricity, we do not expect that including Kozai cycles will significantly alter the estimates of collisional lifetimes. However, they may somewhat alter the behavior of the evolution of inclination and eccentricity over the age of the solar system. Future work is merited to fold in the effects of mutual tides and Kozai cycles, especially when large samples of binaries are uncovered by future surveys and the conclusions derived from their orbital distribution become more precise.

\subsection{Prospects for detecting catastrophic collisions}

Many of the collisions modeled in the numerical simulations we present in \S3 would produce a prodigious amount of dust and debris. Such collisions may occasionally be detectable as transient brightening events, similar to the event detected in the Main Asteroid Belt in early 2010 (Jewitt et al. 2010). Given the frequent deep observations of large areas of the sky in upcoming surveys like Pan-STARRS and LSST, we estimate the frequency at which such events will be detectable. Similar estimates have been made for collisions in the Main Asteroid Belt (LSST Science Book v.2, 2009, Ch. 5.6.1)

Using the estimates of debris cross-sectional area produced in collisions derived in Wyatt \& Dent (2002) for grains larger than 1 mm in radius, and conservatively assuming an albedo of 5\% for the debris produced by the catastrophic disruption of small TNOs, we find that that a collision which disrupts 50\% of the mass of a 700 m radius TNO will produce a debris cloud sufficiently large to be detected in reflected light by LSST (given a single-visit $r$-band depth of 24.7). Similarly, impacts which disrupt 80\% of the mass of a 600 m radius TNO or 5\% of the mass of a 1.6 km radius TNO will also be detectable. By Eqns. \ref{M_lrf} and \ref{Q_star}, we estimate that such disruption events will require impactors of radius $\sim$58 m, 57 m, and 72 m, respectively (given a relative velocity of $\sim 1$ km s$^{-1}$).

The lifetime of these events also factors into their detectability, and depends on the velocity dispersion of the debris cloud. Again adopting the derivations of Wyatt \& Dent (2002) for the rate of azimuthal spreading for debris clouds produced by such a catastrophic impact, we estimate that such systems will remain visible for roughly two to three weeks before their surface brightness drops to below detectable levels.

\begin{figure}
\begin{centering}
\includegraphics[width=0.5\textwidth]{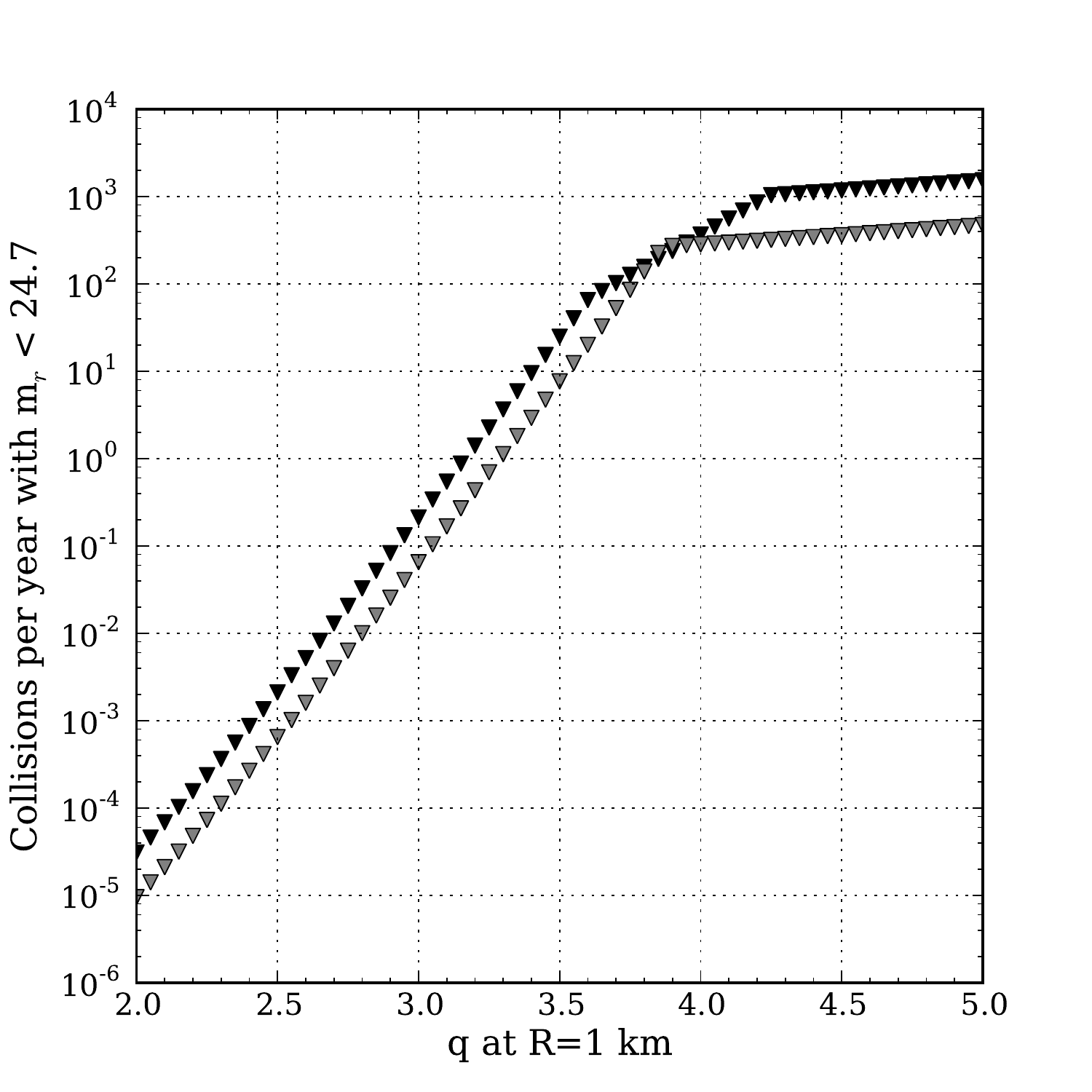}
\caption{Estimate of the rate of collision events in the Classical Kuiper Belt detectable by LSST, given the small-object population limits discussed in the text. Results assuming $P_{i}=1.3\times10^{-21}$ km$^{-2}$ yr$^{-1}$ (black triangles) and $P_{i}=4\times10^{-22}$ km$^{-2}$ yr$^{-1}$ (gray triangles). Breaks indicate transition from population limits determined by the observed number of large objects (leftmost trend), binary survival (middle trend, only visible in the black triangles), and occultation limits (rightmost trend).}
\label{debris}
\end{centering}
\end{figure}

Using the combination of $R\sim1$ km population estimates presented earlier (extrapolation from large size, occultation limits, and binary survival), we can estimate the frequency of impacts as large or larger than this minimum detectable size as a function of size distribution slope. Assuming the same two values for intrinsic collision probability $P_i$ as used in our earlier analysis, we find that for reasonable size distribution slopes, there may be tens to hundreds of collision events detectable by LSST per year. Figure \ref{debris} illustrates the estimated frequency of detectable impacts, and the expected rate of events detectable by LSST is  strongly dependent on the size distribution slope at small size. Thus, the detection or non-detection of these transient collision events in surveys like LSST may prove to be a strong indicator of the size distribution of very small objects in the Kuiper Belt.

\section{Summary}

\begin{enumerate}

\item For most reasonable impactor populations, the collisional lifetime of a TNB can be accurately estimated by analytical arguments with a small empirical correction determined by our simulations; the expression for collisional lifetime is given by Eqn. \ref{final_tau}. This estimate includes the effects of multiple collisions and mass loss.

\item Evolution of separation and eccentricity preferentially occurs along lines of constant apocenter, and 90-95\% of all systems modeled have final pericenters lower than their primordial apocenter. This is further evidence that the ultra-wide binaries are not examples of primordially-tight binaries which have been widened by collisional processes, as we would expect eccentricities to be high on average ($e \gtrsim 0.8$).

\item Collisions with objects in the 1---5 km radius range are capable of unbinding the ultra-wide TNBs, and the continued existence of these systems constrains the number of impactors that can presently exist in the Classical Kuiper Belt. These limits are compatible with the extrapolation of the measured large object ($R >30$ km) population to $R \sim 1$ km with a size distribution power-law slope of less than $q\simeq 3.5-3.7$, depending on the assumed intrinsic collisional probability. These limits are also compatible with the putative detection of a stellar occultation by a single $\sim250$ m object in the Classical Kuiper belt (Schlichting et al. 2009) for slopes greater than $q\simeq2.5-3$. The convergence of these estimates suggests that, barring more complicated structure in the size distribution (eg., a collisional ``divot,'' Fraser 2009), the size distribution slope at small radii is roughly consistent with collisional equilibrium at $q\sim3.5$.

\item Collisions with realistic collider size distributions do not cause any strong asymmetry between prograde and retrograde survival times, and it is likely that the equal numbers of prograde and retrograde mutual orbits reflects the primordial inclination distribution. A non-negligible fraction of binaries have their orientation flipped from prograde to retrograde and vice-versa, but not a large enough fraction to account for the observed prograde to retrograde ratio if the binaries were formed by the $L_2 s$ mechanism, as Schlichting \& Sari (2008) predict roughly 97\% of such systems would form with retrograde orientation.

\item Even with impact trajectories drawn from a disk-like distribution, it is unlikely for the ultra-wide TNB inclinations to have been generated by collisional modification of an initially uniform inclination distribution. Instead, the ultra-wide TNB inclination distribution must have been dynamically colder in the past. 

\item Faster erosion of widely separated binary population plausibly resolves the over-production of ultra-wide binaries by the model of Nesvorn\'{y} et al. (2010), as reasonable impactor populations can easily cause a reduction in the relative fraction of wide binaries to tight binaries by the required factor of roughly four.

\item Analytical arguments can reproduce similar trends in binary fraction with primary radius as found by Nesvorn\'{y} et al. (2011), and slow erosion of the binary population was found to produce similar trends in binary fraction with radius as rapid collisional grinding. This will complicate the interpretation of any future detection of a trend in binary fraction with radius.

\item Upper limits on small-object populations still can allow enough collisions to be occurring that next-generation optical surveys like LSST may detect tens to hundreds of transient brightening events per year due to large dust-producing impacts. The rate of these events is extremely sensitive to the size distribution at small size, and the detection or non-detection of such collisions may be a powerful diagnostic of the decameter-scale impactor population in the current Kuiper Belt.

\end{enumerate}

\section{Acknowledgements}

Alex Parker is funded by the NSF-GRFP award DGE-0836694. We would like to acknowledge the constructive conversations with Jean-Marc Petit and Wesley Fraser regarding the details of our collisional code, and with Mark Booth with respect to dust generation in collisions. We would also like to thank Stephen Gwyn and John Ouellette for their assistance with the computational challenges of this work.

\clearpage

\begin{table}[h]
\centering
\begin{tabular}{lccccccc}
\multicolumn{8}{c}{Table 1}\\
\multicolumn{8}{c}{Fit parameters for Eqn. \ref{tau0} \& adopted parameters for Eqn. \ref{final_tau}}\\
\hline
Name & K (yr) & $r_c$ (km)& $r_i$$^{a}$ (km) & $M_{sys}$ (kg) & $R_p^b$ (km) & $R_s^b$ (km) & $a_m$ (km) \\
\hline
\hline
2000 CF$_{105}$   & 3.81$\times10^7$ & 1.60 & 1.65 & 1.85$\times10^{17}$ & 32.0 & 23.0 & 3.33$\times10^{4}$ \\
2001 QW$_{322}$ & 2.54$\times10^6$ & 4.50 & 5.74 & 21.1$\times10^{17}$ & 63.5 & 63.5 & 1.015$\times10^{5}$ \\
2003 UN$_{284}$  & 4.50$\times10^6$ & 3.79 & 3.83 & 12.4$\times10^{17}$ & 62.2 & 41.6 & 5.55$\times10^{4}$ \\
2005 EO$_{304}$  & 3.43$\times10^6$ & 4.31 & 3.74 & 20.7$\times10^{17}$ & 76.2 & 39.1 & 6.98$\times10^{4}$ \\
2006 BR$_{284}$                      & 8.38$\times10^6$ & 3.08 & 3.26 & 5.7$\times10^{17}$ & 44.9 & 35.7 & 2.53$\times10^{4}$ \\
2006 JZ$_{81}$                        & 4.33$\times10^6$ & 4.00 & 3.83 & 12.1$\times10^{17}$ & 60.8 & 38.7 & 3.23$\times10^{4}$ \\
2006 CH$_{69}$                       & 4.46$\times10^6$ & 3.92 & 3.95 & 8.4$\times10^{17}$ & 50.4 & 41.2 & 2.76$\times10^{4}$ \\
\hline
\end{tabular}

\footnotesize{ $^{a}$: Required impactor radius for disruption by collision with secondary, from Eqn. \ref{R_est}. \\
$^{b}$: Radii of components are estimated by adopting system mass and delta-magnitudes measured in Parker et al. (2011) and assuming a bulk density of 1 gram cm$^{-3}$.}
\end{table}

\begin{figure}
\begin{centering}
\includegraphics[width=0.5\textwidth]{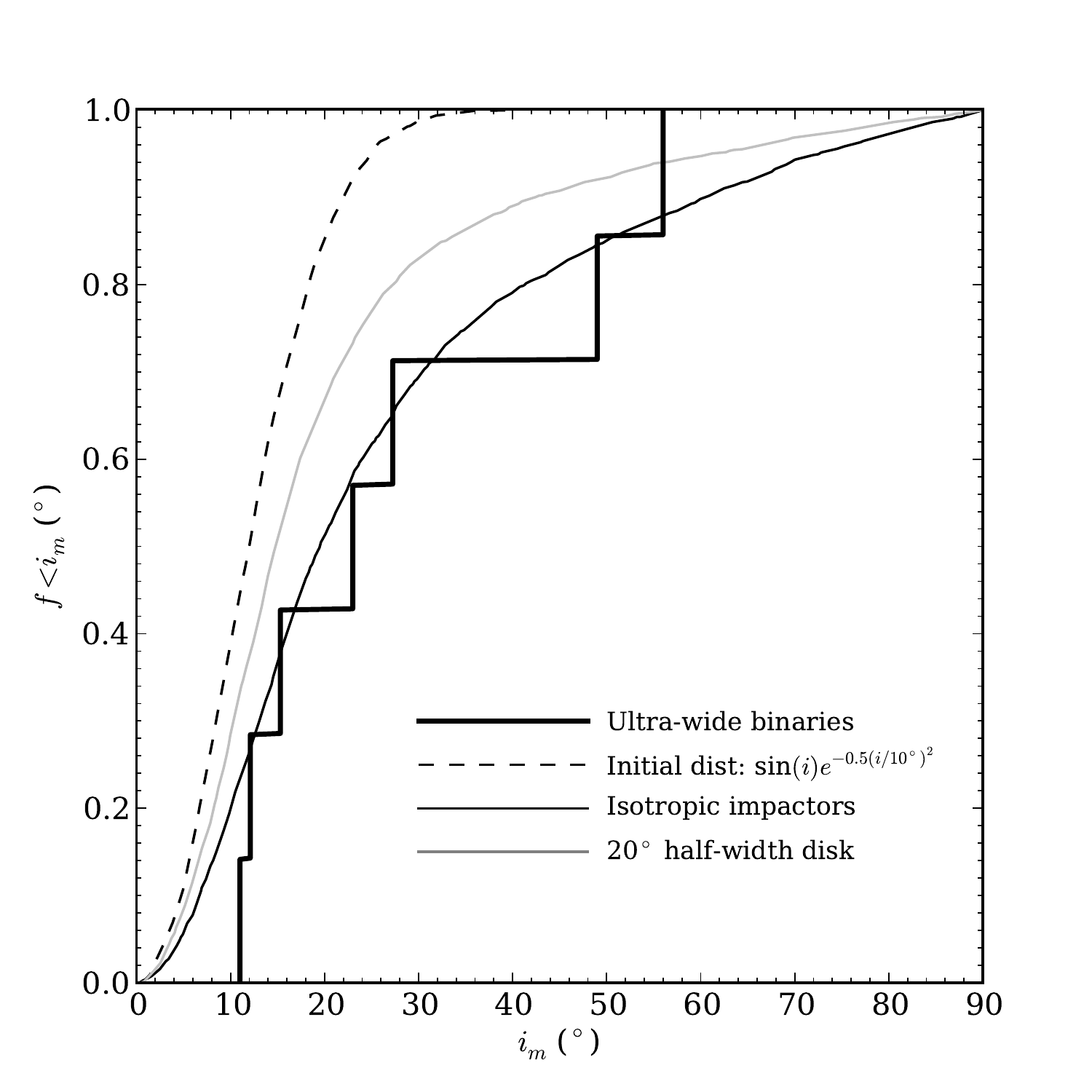}
\caption{Comparison of initial and final inclination distributions after collisional evolution (final inclination taken just prior to binary disruption). Initial inclinations drawn from $p(i) \propto \sin(i)e^{-\frac{1}{2}\left( \frac{i}{10^{\circ}} \right)^2}$ (dashed line). Final inclinations shown for random impact trajectories (light black line) and disk-like impactor geometry (light gray line), and current ultra-wide TNB inclinations shown for comparison (heavy black line). }
\label{collide_dist}
\end{centering}
\end{figure}

\end{document}